\documentclass[twocolumn]{aastex62}
\usepackage{multirow}
\usepackage{pifont}

\def \nustar {{\em NuSTAR}}
\def \smc {{SMC~X-1}}
\def \companion {{Sk~160}}
\def \caltech {{Cahill Center for Astronomy and Astrophysics, California Institute of Technology, Pasadena, CA 91125, USA}}

\newcommand{\code}{\texttt}
\newcommand{\comment}[1]{}

\shorttitle{Transient Pulsations in \smc}
\shortauthors{Pike et al.}

\begin{document}
\title{Observing the transient pulsations of \smc\ with \nustar}
\author{Sean N. Pike}
\author{Fiona A. Harrison}
\affil{\caltech}
\author{Matteo Bachetti}
\affil{Osservatorio Astronomico di Cagliari, Cagliari, Italy}
\author{McKinley C. Brumback}
\affil{Department of Physics and Astronomy, Dartmouth College, Hanover, NH 03755, USA}
\author{Felix S. F\"urst}
\affil{European Space Astronomy Centre (ESAC), Science Operations Department, Madrid, Spain}
\author{Kristin K. Madsen}
\affil{\caltech}
\author{Katja Pottschmidt}
\affil{CRESST, Department of Physics, and Center for Space Science and Technology, UMBC, Baltimore, MD, 21250, USA}
\affil{NASA Goddard Space Flight Center, Greenbelt, MD, 20771, USA}
\author{John A. Tomsick}
\affil{Space Sciences Laboratory, University of California, Berkeley, CA 94720, USA}
\author{J\"orn Wilms}
\affil{Astronomical Institute of the University of Erlangen-Nuremberg, Bamberg, Germany}

\begin{abstract}
We report on \nustar\ observations of transient pulsations in the neutron star X-ray binary \smc. The transition from non-pulsing to pulsing states was not accompanied by a large change in flux. Instead, both pulsing and non-pulsing states were observed in a single observation during the low-flux super-orbital state. During the high-state, we measure a pulse period of $P = 0.70117(9)\,\mathrm{s}$ at $T_{ref} = 56145\,\mathrm{MJD}$. Spectral analysis during non-pulsing and pulsing states reveals that the observations can be consistently modeled by an absorbed power law with a phenomenological cutoff resembling a Fermi-Dirac distribution, or by a partially obscured cutoff power law. The shapes of the underlying continua show little variability between epochs, while the covering fraction and column density vary between super-orbital states. The strength of pulsations also varies, leading us to infer that the absence and reemergence of pulsations are related to changing obscuration, such as by a warped accretion disk. \smc\ is accreting near or above its Eddington limit, reaching an unabsorbed X-ray luminosity of $L_{\rm X}({\rm  2-10~keV}) \approx 5 \times 10^{38}\, {\rm erg}\, {\rm s}^{-1}$. This suggests that \smc\ may be a useful local analog to ultraluminous X-ray pulsars (ULXPs), which likewise exhibit strong variability in their pulsed fractions, as well as flux variability on similar timescales. In particular, the gradual pulse turn-on which has been observed in M82~X-2 is similar to the behavior we observe in \smc. Thus we propose that pulse fraction variability of ULXPs may also be due to variable obscuration.
\end{abstract}


\section{Introduction}
\label{sec:intro}
High Mass X-ray binary (HMXB) systems, in which a compact object accretes matter from a more massive companion, often exhibit variability on multiple timescales ranging from less than a second to several months. For most HMXBs, accretion onto the compact object is fed by the stellar winds of the companion. However, in some cases, the objects may orbit at a small enough distance that the companion fills its Roche lobe, resulting in higher accretion rates and higher luminosities. In the case of accreting neutron stars, matter is funnelled along the magnetic field lines onto the surface of the star, resulting in a column of accreted material at the magnetic poles. Because the magnetic poles and the spin axes are not perfectly aligned, the accretion column revolves at the same rate as the neutron star, resulting in emission which appears pulsed.

One such HMXB which has been well-studied is \smc. This system, residing in the Small Magellanic Cloud, was first detected by \citet{Price1971}. It was later resolved as a discrete source by \citet{Leong1971}, who reported significant variability in both the intensity and spectrum of the source. The binary nature of \smc\ was soon confirmed by \citet{Schreier1972} who discovered periodic occultations with an orbital period of around 3.9 days. \smc\ also exhibits pulsations with a period of about 0.7 seconds, and the pulse fraction and shape are known to vary significantly over time \citep{Lucke1976}. The existence of X-ray pulsations confirms that the accreting compact object is a neutron star. Accretion onto the neutron star has been attributed to Roche lobe overflow \citep{Hutchings1977,VanParadijs1977} of its companion, \companion, which has been spectrographically classified as a B0 I supergiant \citep{Webster1972}. This classification places \smc\ in a subcategory of HMXBs known as supergiant X-ray binaries (SGXB). Finally, the source exhibits super-orbital variability on a timescale of 45 to 60 days, which has been attributed to obscuration by a precessing tilted accretion disk \citep{Wojdowski1998}. Throughout this paper, we assume a distance to \smc\ of 60.6\,kpc as reported by \citet{Hilditch2005}.

\begin{figure*}[htp]
	\gridline{\fig{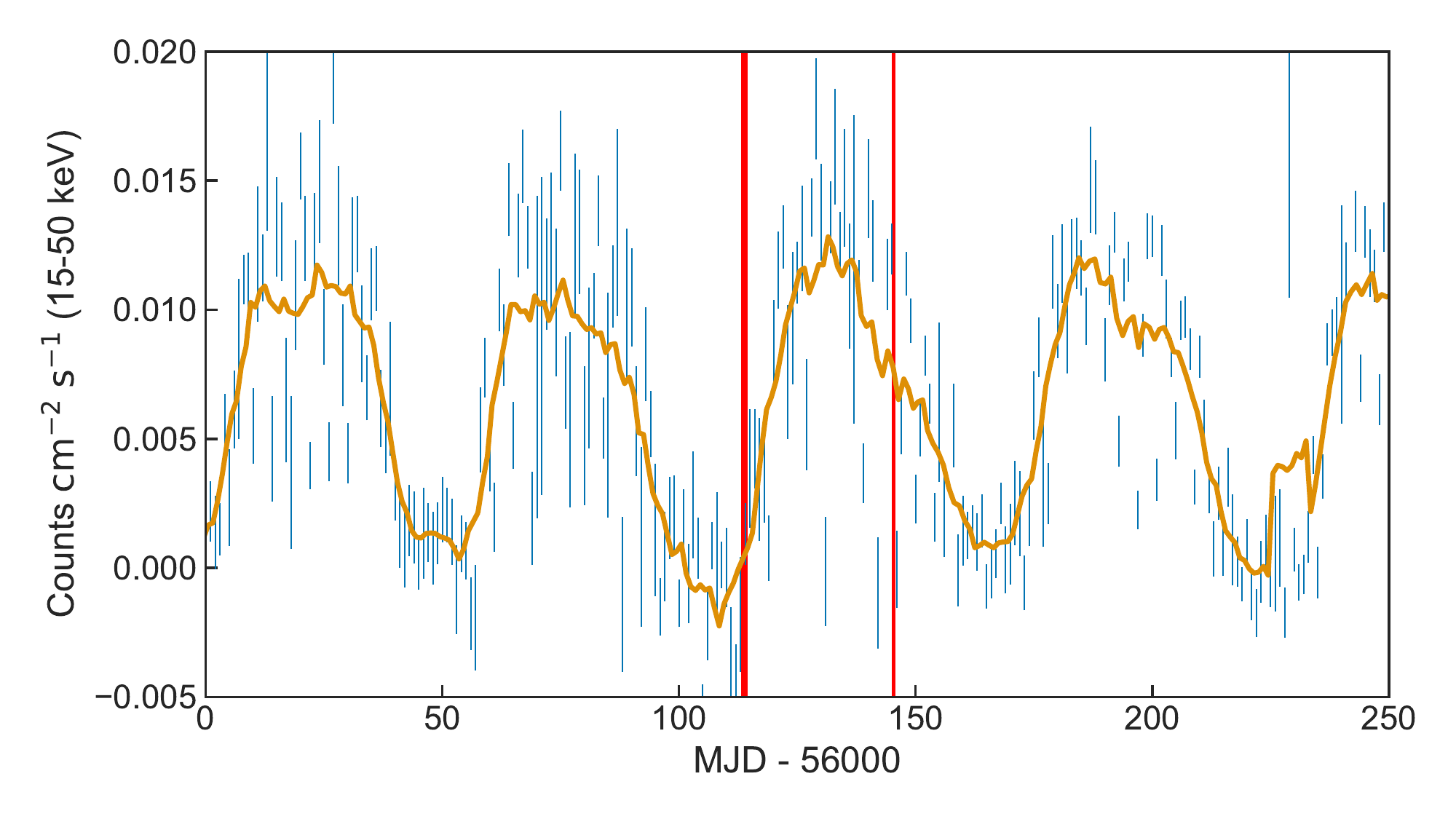}{0.7\textwidth}{}}
	\caption{{\em Swift} BAT light curve of \smc\ during 2012. The moving average of the BAT flux is shown in gold. A super-orbital period of around $60$\,days is clearly visible. The red vertical bars indicate the duration of each \nustar\ observation presented here. The first observation (10002013001) took place near the end of the low state, while the second observation (10002013003) took place as the source was growing fainter shortly after the high state.}
	\label{fig:BATcurve}
\end{figure*}

\begin{figure*}[htp]
	\gridline{\fig{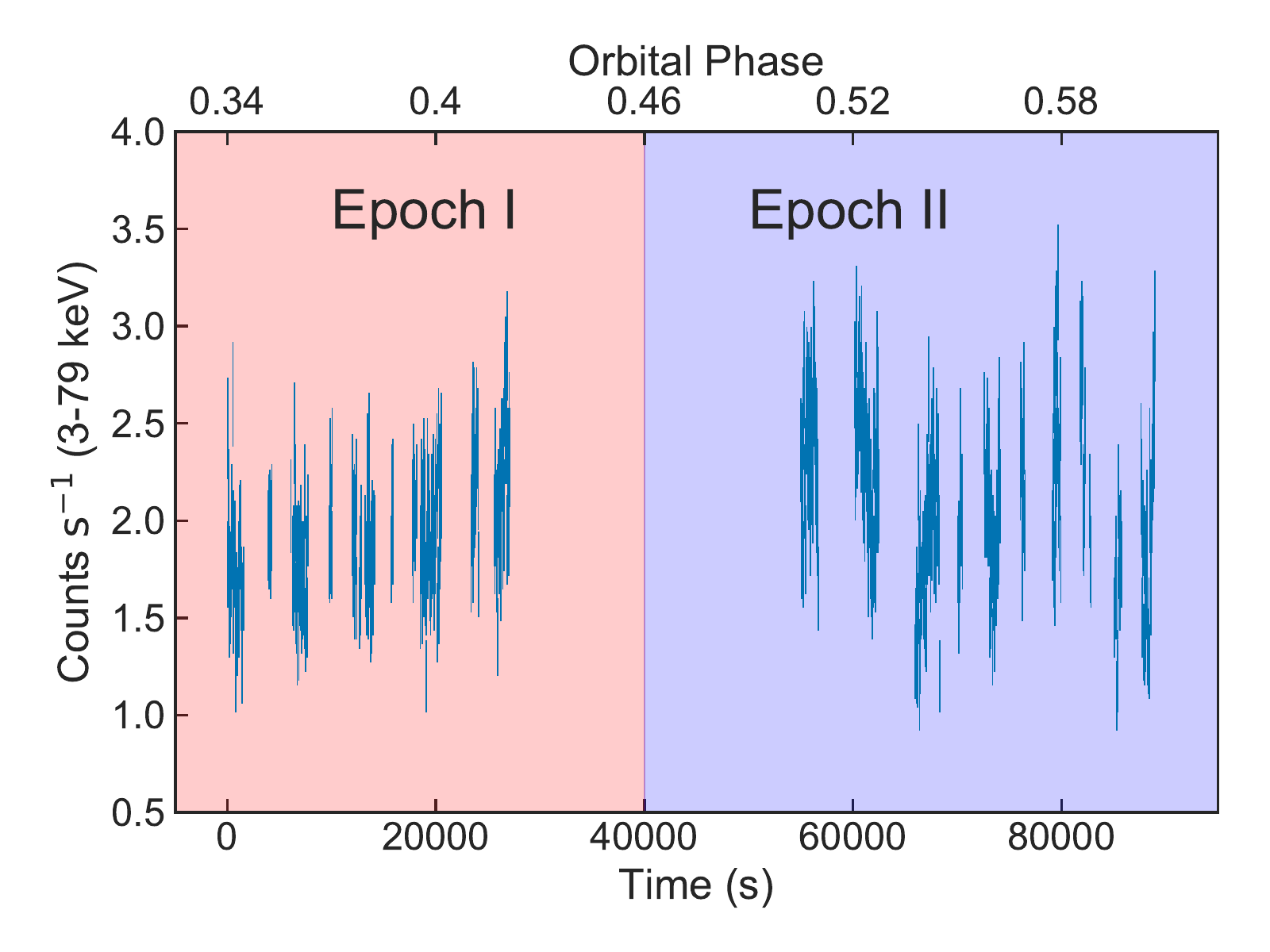}{0.49\textwidth}{(a)}
	\fig{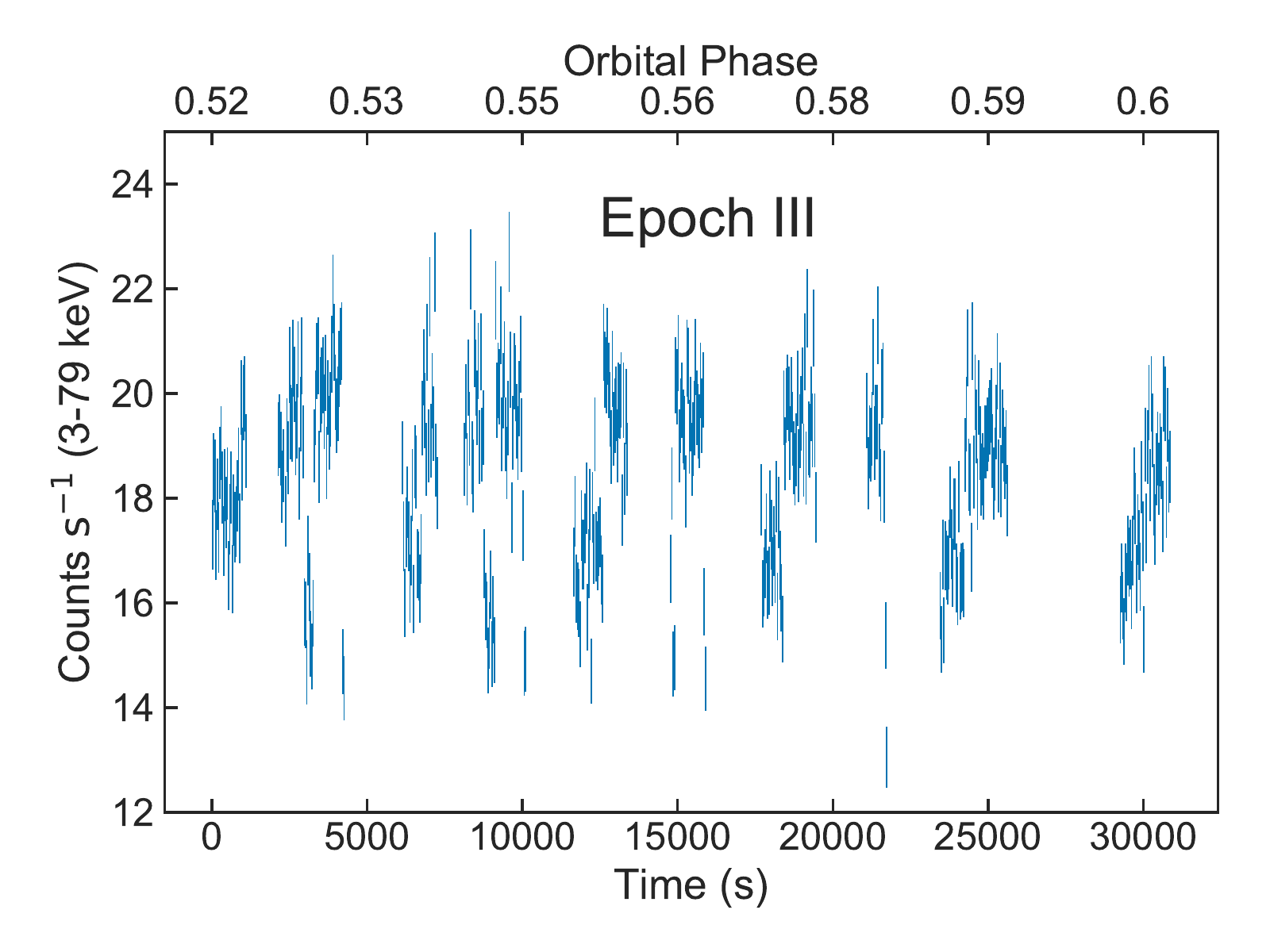}{0.49\textwidth}{(b)}}
	\caption{(a) \nustar\ FPMA count rate of the source during the first observation (10002013001). The gap between 30000 and 50000 seconds is due to a failed data downlink and is not inherent to the source. We have split this observation into two epochs. (b)  \nustar\ FPMA count rate of the source during the second observation (10002013003). We define the third epoch as the entirety of this observation. The apparent variability on timescales of $\sim5000$\,s during Epoch III can be attributed to movement of the source between detectors. Both light curves are binned into intervals of 40\,s. The orbital phase, which is defined by full eclipse of the source at $\phi=0$ and $\phi=1$, is included along the horizontal axis as well as the time in seconds since the beginning of the each observation.}
	\label{fig:lightcurves}
\end{figure*}

One of only a handful of SGXBs known to accrete via Roche lobe overflow, \smc\ exhibits persistent emission near or above its isotropic Eddington luminosity of $L_{\rm Edd}\sim 1.3\times10^{38}\,\mathrm{erg\,s^{-1}}$ (for a mass of $\sim 1.1\, M_{\odot}$, as reported by \citealt{VanderMeer2007}) varying between $L_{\rm X}(2-12\,{\rm keV}) \sim 10^{37}\ \mathrm{erg}\ \mathrm{s}^{-1}$ in the low state and luminosities in excess of $5\times10^{38}\ \mathrm{erg}\ \mathrm{s}^{-1}(2-12\,{\rm keV})$, more than three times its Eddington luminosity, in the high state \citep{Bonnet-Bidaud1981}. In addition to this persistent emission, \smc\ has been shown to exhibit type II X-ray bursts with durations of tens of seconds \citep{Angelini1991THEL,2018RAA....18..148R}. Its near- to super-Eddington luminosity places the source in a middle ground between less luminous Be/X-ray binaries (BeXB), which exhibit a range of persistent X-ray luminosities from $10^{32}\ \mathrm{erg}\ \mathrm{s}^{-1}$ \citep{Tomsick2011} up to $10^{35}\ \mathrm{erg}\ \mathrm{s}^{-1}$ \citep{Reig1999}, and brighter ultraluminous X-ray pulsars (ULXPs).

ULXPs, the known examples of which are M82~X-2 \citep{Bachetti2014},  NGC~7793~P13 \citep{Furst2016, Israel2017}, NGC~300~ULX1 (also SN~2010da, \citeauthor{Carpano2018DiscoveryEvolution} \citeyear{Carpano2018DiscoveryEvolution}) and NGC~5907~ULX-1 \citep{Israel2017a}, vary between bright pulsing states during which the luminosity can reach $10^{41}\ \mathrm{erg}\ \mathrm{s}^{-1}$ --- several orders of magnitude above the Eddington limit of a typical neutron star --- and faint states when the luminosity drops to $10^{37-38}\ \mathrm{erg}\ \mathrm{s}^{-1}$ \citep{Kaaret2017}. Similar to \smc\ \citep[]{Inam2010}, these sources exhibit pulsations with periods on the order of one second and spin-up rates of $|\dot{P}|=10^{-11}-10^{-9}\,\mathrm{s\,s^{-1}}$, with the exception of NGC~300~ULX1, which has a much longer pulse period of 32\,s and a faster spin-up rate on the order of $10^{-7}\,\mathrm{s\,s^{-1}}$. 

A given ULXP may not exhibit detectable pulsations at all times, and when they are detected, the fraction of their flux which is pulsed is variable. Pulse transience has been attributed to the propeller effect in which rotation of the neutron star's magnetosphere halts accretion by flinging accreting material out of the system before it can reach the corotation radius \citep{Illarionov1975}. In contrast to the flux variability of \smc\ which occurs quasi-periodically with a continuous transition between high and low states, the propeller effect is associated with changes of more than a factor of $40$ in luminosity on shorter timescales which results in a bimodal flux distribution in ULXPs \citep{Tsygankov2016}. In terms of pulse fraction, this bimodality corresponds to distinct pulsed and non-pulsed states. However, continuous variability in pulse fraction has also been observed in ULXPs. In particular, the pulse fraction of M82~X-2 was shown to gradually increase from $8$\% to $23$\% in the $10-30$\,keV range over an interval of around 10 days \citep{Bachetti2014}.

Periodic variability on timescales of 60-80 days has also been measured for the ULXPs NGC 7793 P13, NGC 5907 ULX-1, and M82 X-2 \citep[][respectively]{Motch2014, Walton2016, 2019arXiv190110491B}. While the ~64 day period observed in NGC~7793~P13 has been attributed to the orbital motion of the binary \citep{Furst2018AP13}, this variability has been classified as super-orbital in the case of M82 X-2. It is still uncertain whether the 78\,day period observed in NGC 5907 ULX-1 is orbital or super-orbital in nature. Given that \smc\ displays super-orbital modulations on similar timescales, as well as its persistent near- to super-Eddington luminosity and variable pulsations, \smc\ may provide a link between ULXPs and classes of X-ray binaries which have been studied in more detail. 

In this paper, we present timing and spectral analyses of two observations of \smc. In Section \ref{sec:data}, we describe the observations of \smc\ and our data reduction methods, including data extraction and corrections. In Section \ref{sec:timing}, we describe the methods and results of our timing analysis, and in Section \ref{sec:spectral}, we describe the spectral analysis of \smc. In Section \ref{sec:discussion}, we discuss the results of our analyses and offer a physical interpretation. Finally, in Section \ref{sec:conclusions}, we list our conclusions and discuss possible applications of our analysis to studies of ULXPs.

\section{Observations and Data Reduction}
\label{sec:data}

\smc\  was observed twice by the {\em Nuclear Spectroscopic Telescope Array} (\nustar) \citep{Harrison2013} in 2012, in the first two months after the launch of the satellite for the purpose of calibration. \nustar\ consists of two focal plane modules, FPMA and FPMB, each of which is made up of four pixelated detectors (DET0-DET3). Each module has a field of view of about 12\,arcminutes, and, combined with focusing optics at a focal length of 10\,m, achieves an angular resolution of 18\,arcseconds, full width at half maximum (FWHM). The energy resolution, given by the FWHM, is 400\,eV at 10\,keV and 900\,eV at 68\,keV, and the full energy range is 3-79\,keV. The timing resolution of the onboard clock is 2\,$\mu$s with a dead time of 2.5\,ms, leading to a maximum count rate of around 400\,events\,$\mathrm{s^{-1}}$.

The first observation took place on 2012 July 5 (OBSID 10002013001) and the second took place on 2012 August 6 (OBSID 10002013003) with exposure times of 27\,ks and 15\,ks, respectively. Figure \ref{fig:BATcurve} shows the light curve of the source as observed by the {\em Neil Gehrels Swift Observatory} Burst Alert Telescope (BAT) during a 250\,day interval bracketing the \nustar\ observations in 2012. The super-orbital period of around 60\,days is clearly visible, and the red bars show the location and duration of each observation in the super-orbital cycle. The first \nustar\ observation occurred at the end of the low state, when the luminosity was just beginning to rise, while the second observation occurred near the end of the high state, when the source was growing fainter. The observations were planned such that they avoided obscuration effects due to the donor star.

We reduced the data using version 1.8.0 of the NuSTARDAS pipeline and \nustar\ CALDB v20170817. We used DS9 \citep{ds9} to select a circular source region with radius 55 arcseconds centered on the position of the source determined by automatic centroid detection. We also selected a circular background region with radius 80 arcseconds located on the same detector as the source, taking care to choose a region free of other sources and outside the source distribution. We corrected the photon arrival times to the solar system barycenter using the position of the source used for data extraction. Before analysis, the photon arrival times were also corrected for the orbital motion of the source using parameters reported by \citet{Falanga2015} and \citet{Inam2010}.

We define three epochs of observation, labeled Epochs I, II, and III. The \nustar\ light curve for each observation is shown in Figure \ref{fig:lightcurves}. Epoch I is defined as the first 40\,ks (13\,ks of exposure time) of observation 10002013001, while the latter half (14\,ks of exposure) of observation 10002013001 makes up Epoch II. The whole of observation 10002013003 makes up Epoch III, which has an exposure time of 15\,ks. During observation 10002013001, the source was positioned on DET0, while the source was positioned on DET3 near the gap between DET3 and DET0 during observation 10002013003. Movement of the source between the two detectors accounts for the $\sim5000$\,s variability apparent in Figure \ref{fig:lightcurves}b. The background count rate did not vary significantly between observations, and for all three epochs, the background rate remained below $10\%$ of the total count rate for energies up to $\sim50$\,keV. To avoid background contamination, we performed spectral analysis for energies between 3\,keV and 40\,keV, resulting in $5.2\times10^4$, $6.3\times10^4$, and $6.3\times10^5$ spectral counts (combined FPMA and FPMB) for Epochs I, II, and III, respectively. For the purpose of spectral analysis, we binned the data such that there are at least 50 counts in each energy bin in Epoch I and Epoch II, and at least 100 counts in each energy bin in Epoch III. We chose to bin Epoch III with more events per bin due to the significantly higher count rate during that epoch.

\section{Timing Analysis}
\label{sec:timing}

\begin{figure*}[htb]
	\gridline{\fig{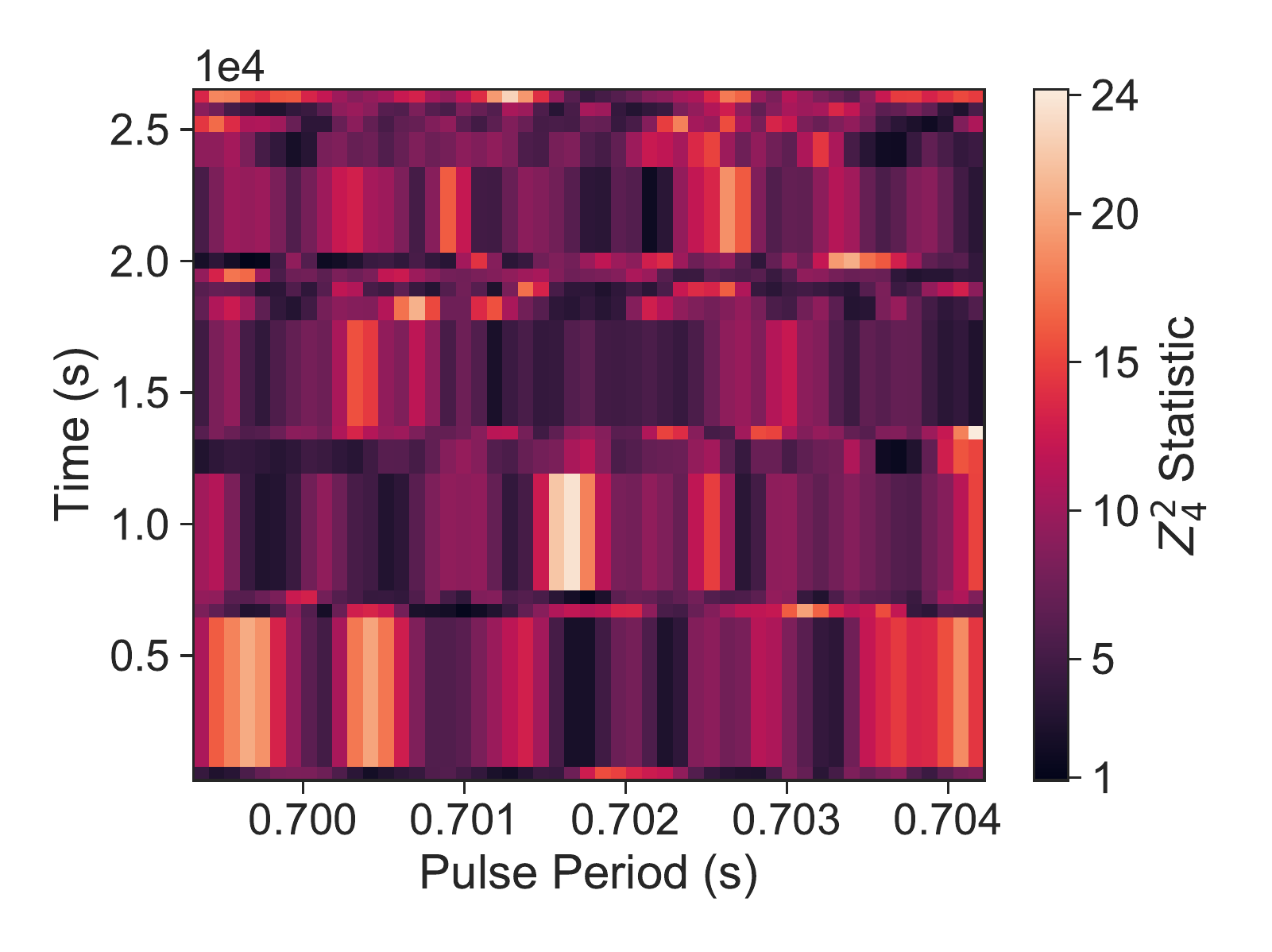}{0.33\textwidth}{(a) Epoch I dynamic folding search}
	\fig{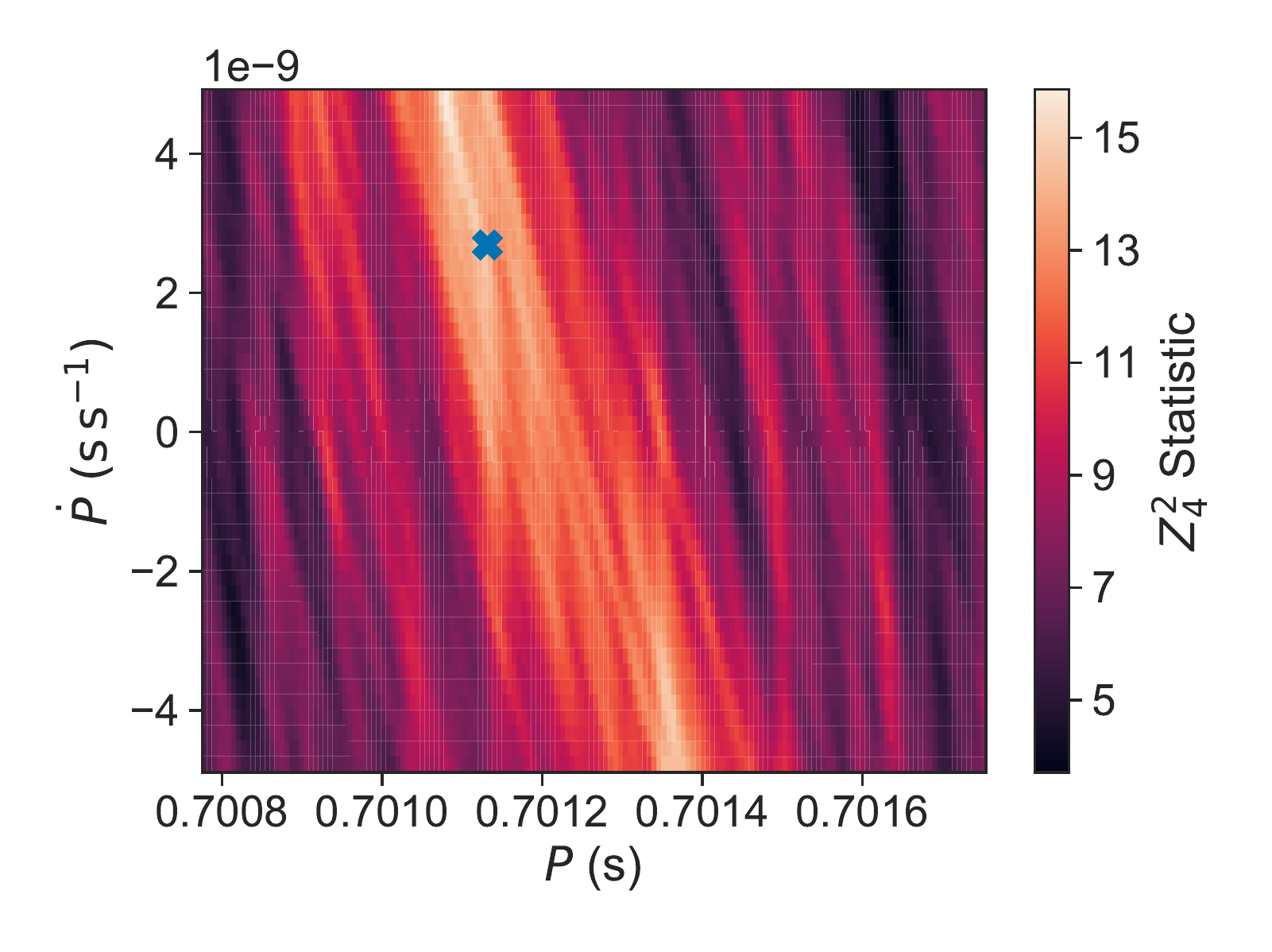}{0.33\textwidth}{(b) Epoch I period search}
	\fig{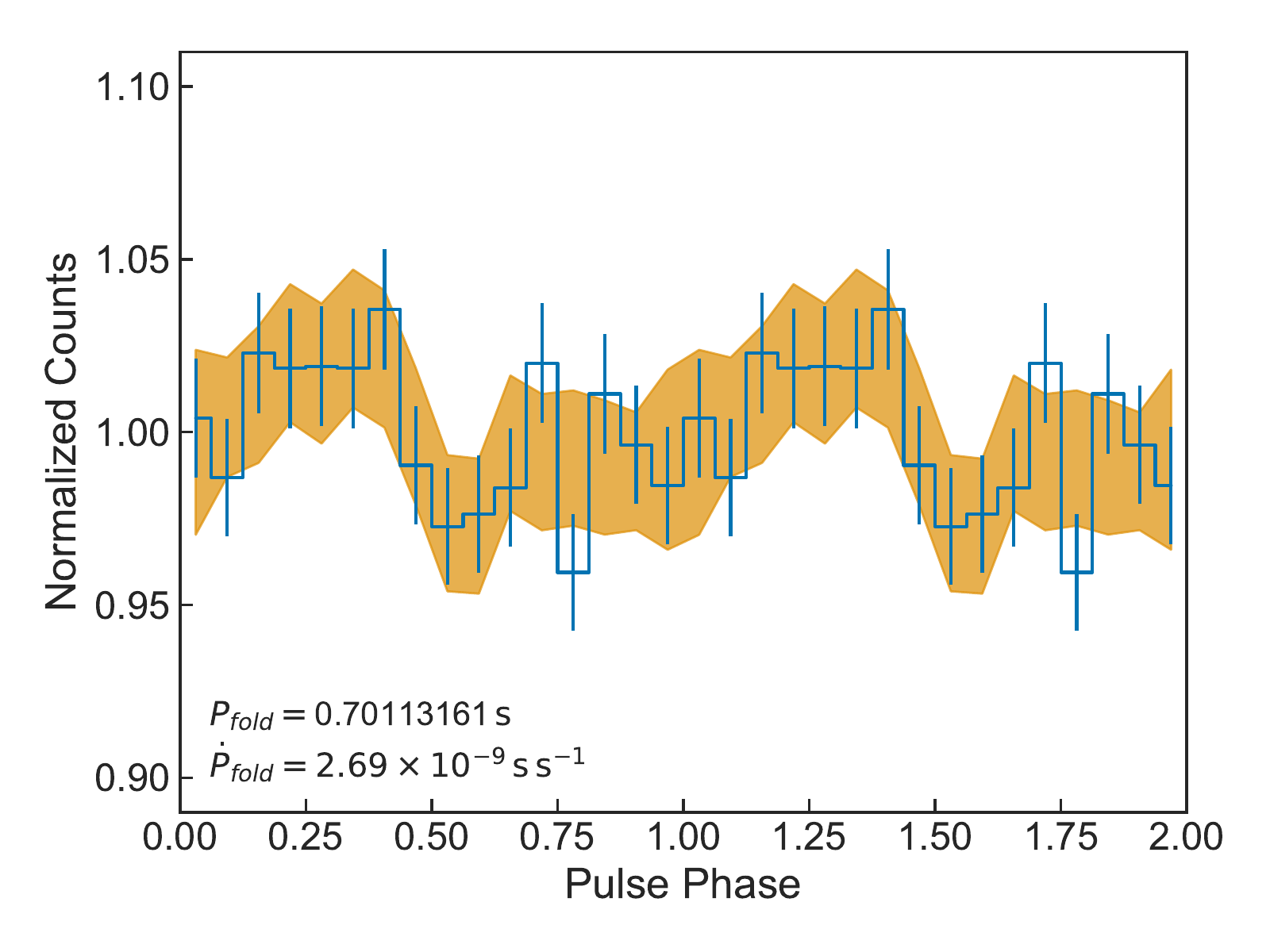}{0.33\textwidth}{(c) Epoch I pulse profile}\label{fig:epochIsearches}}
	\gridline{\fig{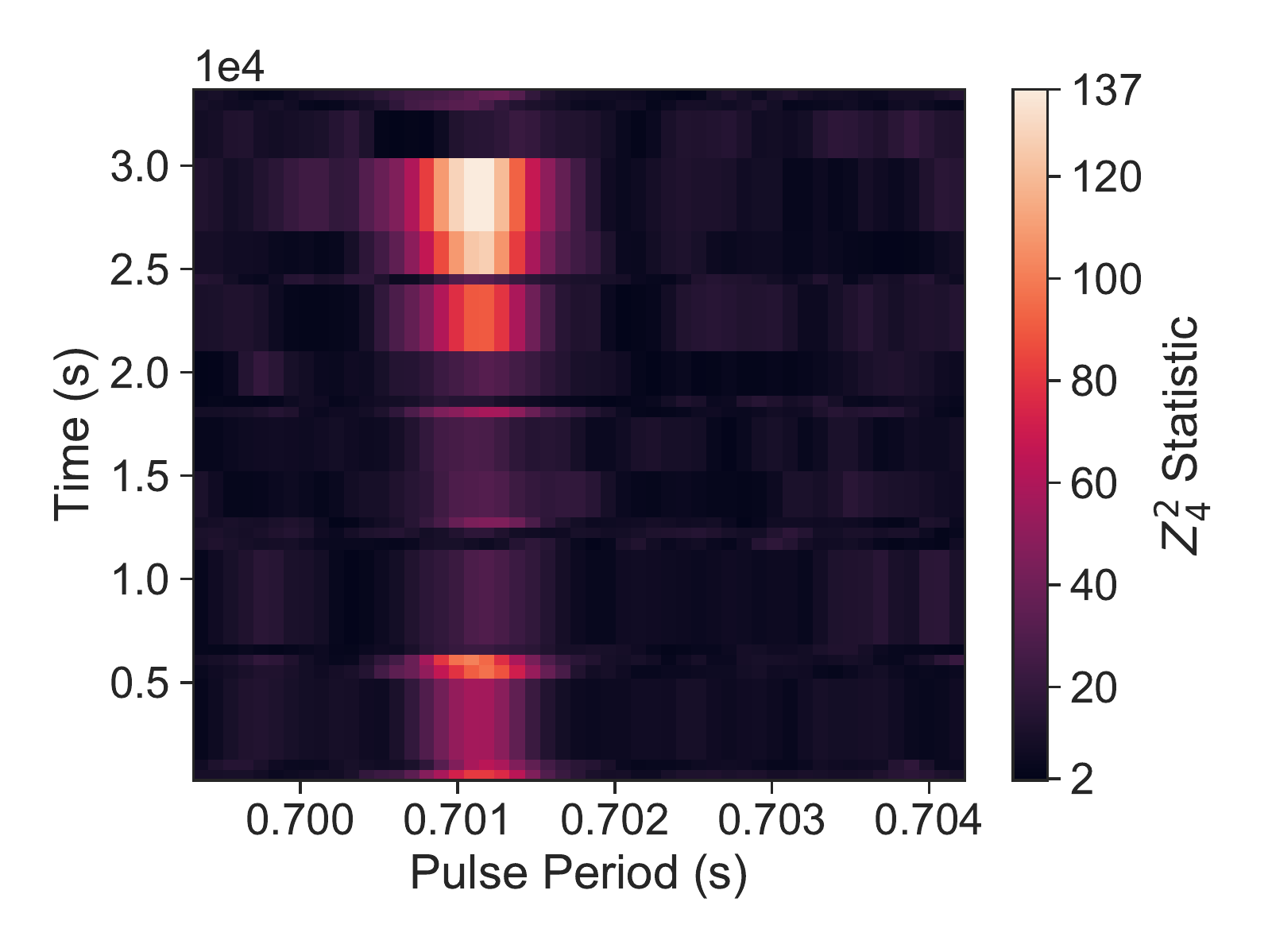}{0.33\textwidth}{(d) Epoch II dynamic folding search}
	\fig{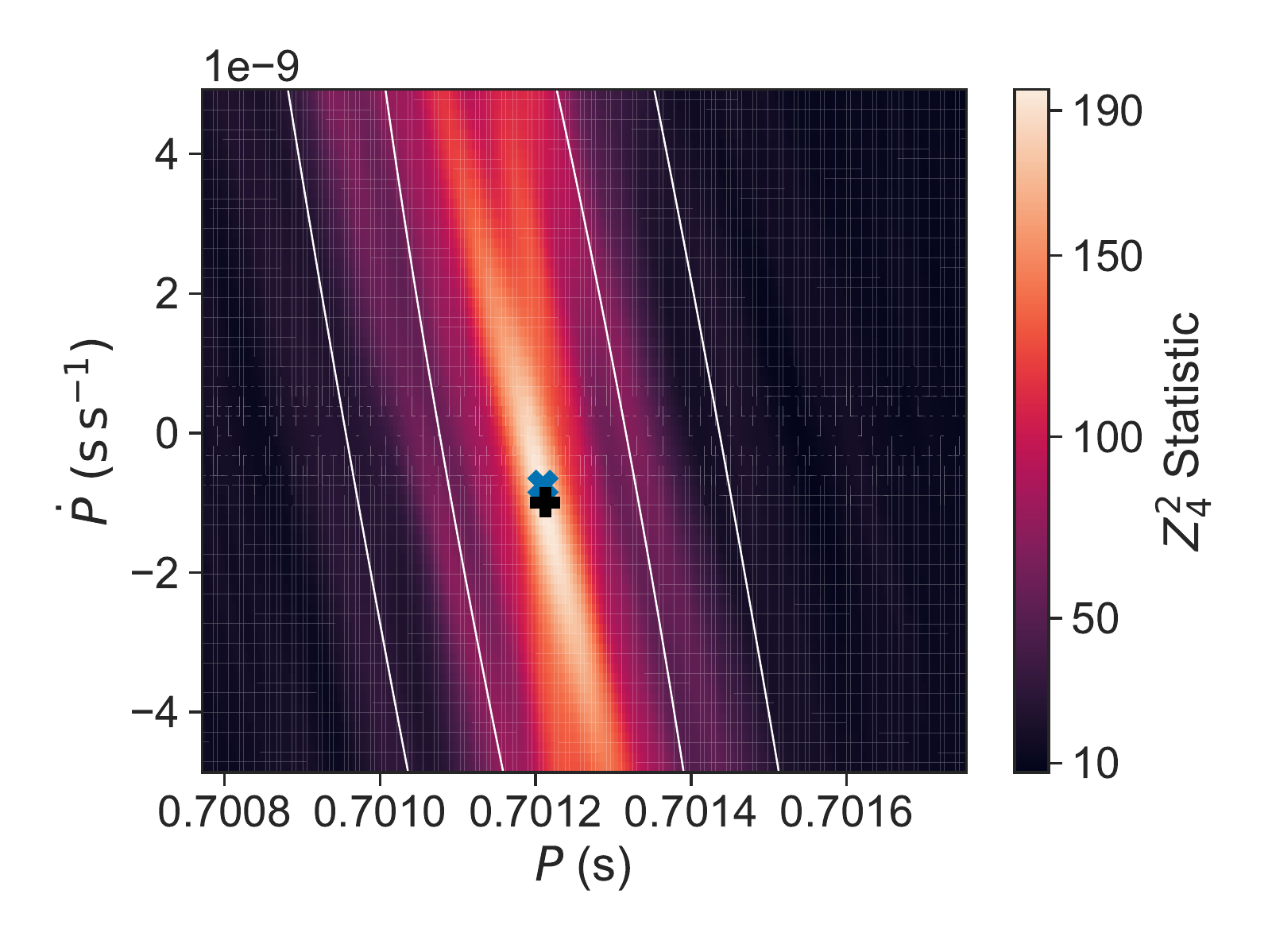}{0.33\textwidth}{(e) Epoch II period search}
	\fig{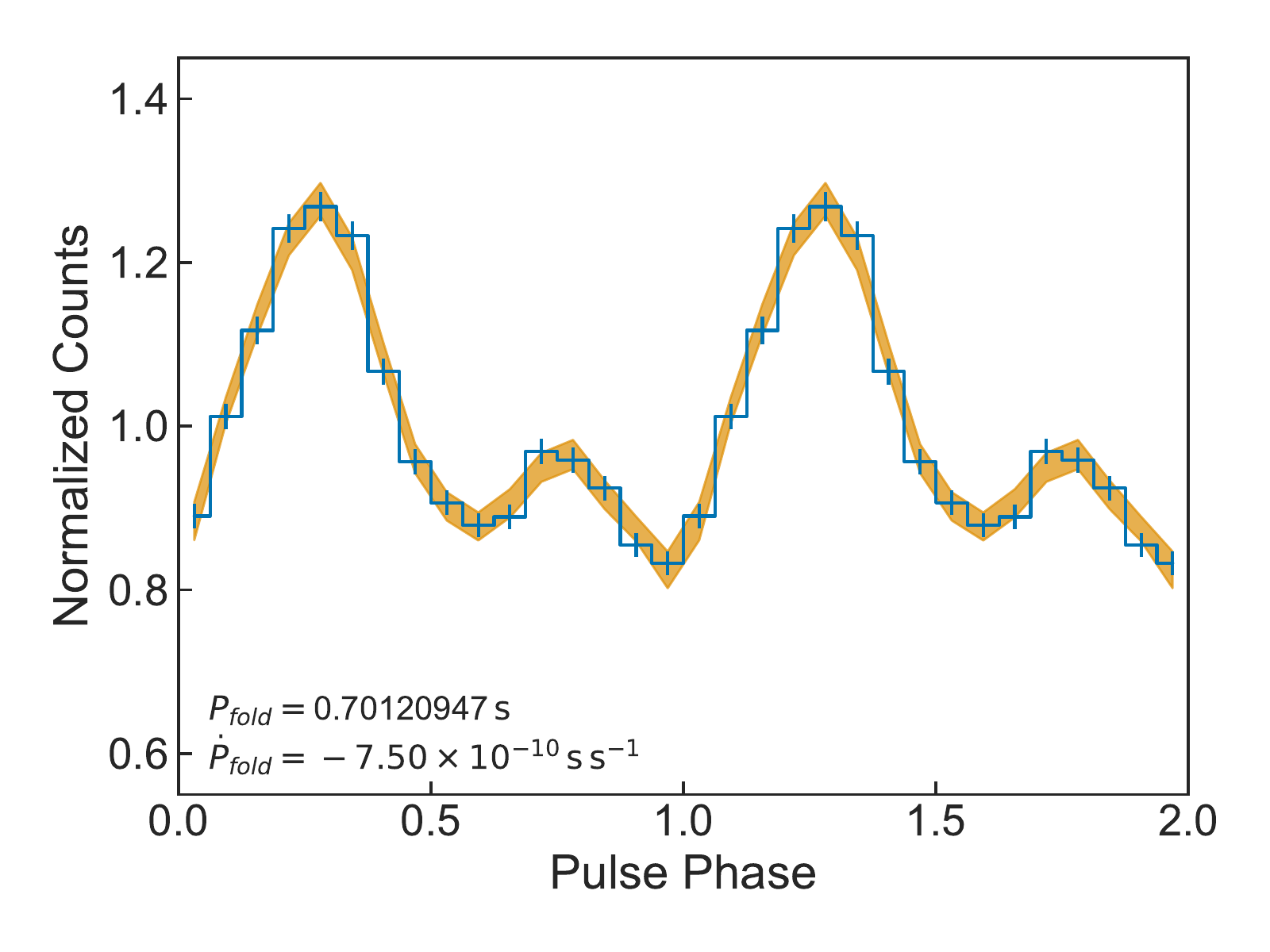}{0.33\textwidth}{(f) Epoch II pulse profile}\label{fig:epochIIsearches}}
	\gridline{\fig{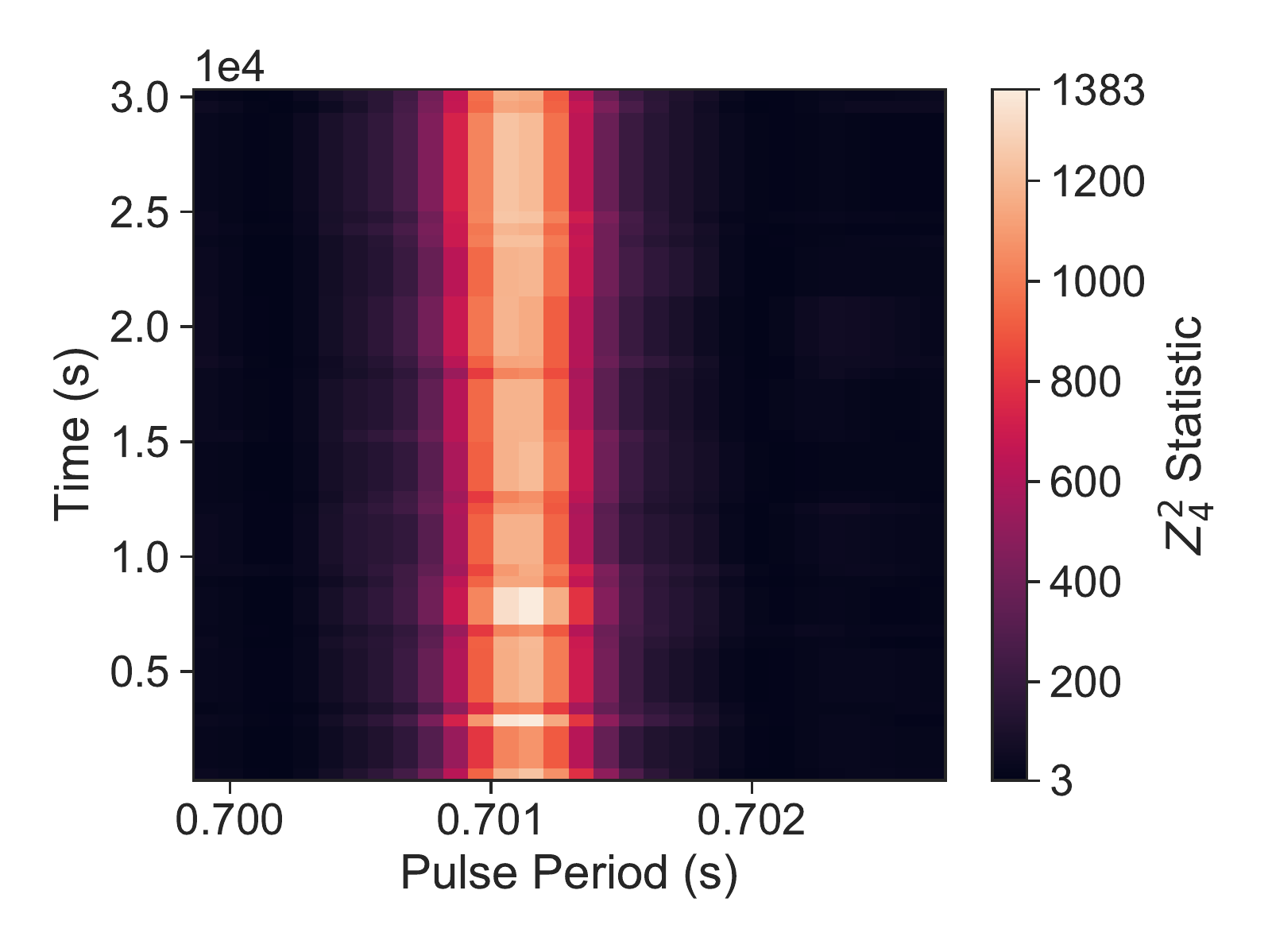}{0.33\textwidth}{(g) Epoch III dynamic folding search}
	\fig{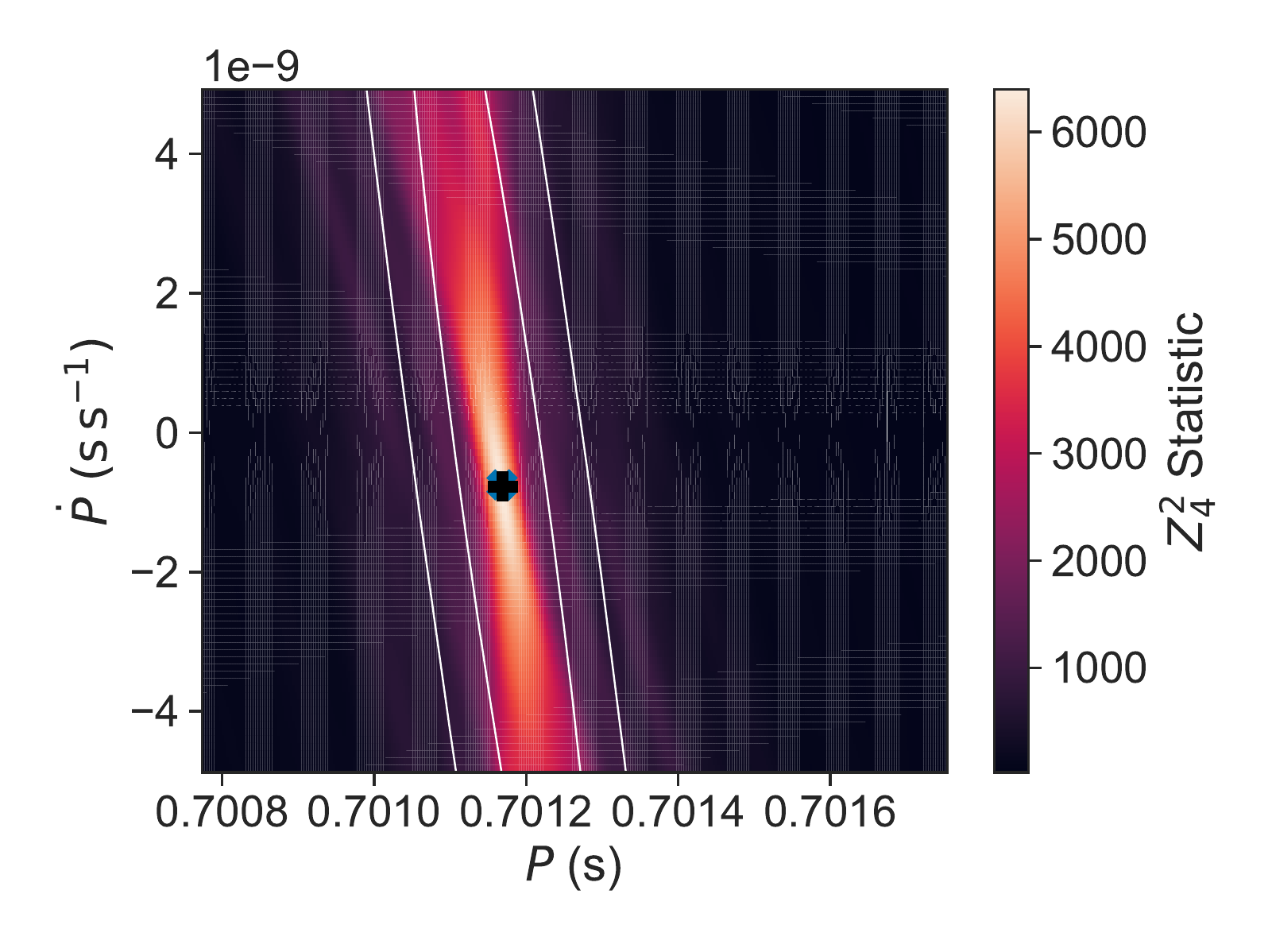}{0.33\textwidth}{(h) Epoch III period search}
	\fig{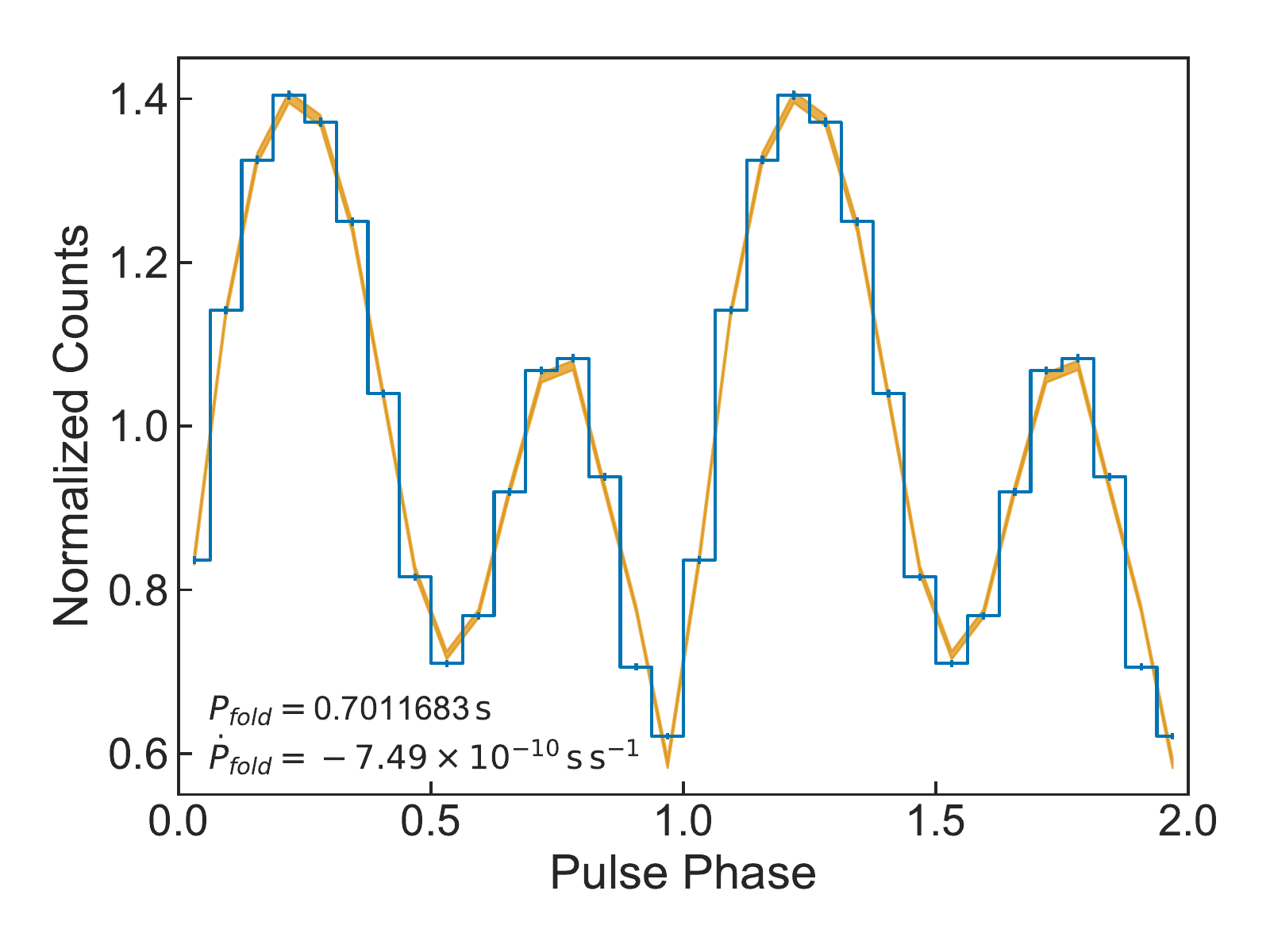}{0.33\textwidth}{(i) Epoch III pulse profile}\label{fig:epochIIIsearches}}
	\caption{Results of pulsation searches applied to each epoch. The left column shows the results of a dynamic folding search. Pulsations are not detected during Epoch I but seem to appear and gradually increase in strength after observation continues during Epoch II. Pulsations are clearly detected in Epoch III and do not appear to vary significantly throughout the epoch. The middle column shows the results of folding searches over both the pulse period and its first derivative. The results of the dynamic searches allowed us to search over a narrower period range. The resulting $Z^2_4$ distribution (b, e, and h) for each epoch is fitted to a 2-d Gaussian distribution. The mean of the fitted Gaussian is indicated by a black cross (\ding{58}) while the white contours represent the 1- and 2-sigma confidence regions. The apparent correlation between $P$ and $\dot{P}$ is an artifact of the search itself and is not intrinsic to the source. The maximum $Z^2_4$ value achieved by each search is indicated by a blue cross (\textcolor{blue}{\ding{54}}) and was used to produce pulse profiles shown in blue in panels b, d, and f. In gold are the 90\% confidence regions determined by the Monte Carlo procedure described in Section \ref{sec:intro}. When applied to Epoch I, the search produces multiple maxima of relatively low detection probability, resulting in a poor fit which cannot constrain the pulse period and first derivative to within the search bounds. We therefore do not show the fitted Gaussian, and we choose to fold the pulse profile using the maximum nearest the values measured for Epochs II and III.  The result is a profile with weak pulsations which are not detected when the last 5000\,s of Epoch I are omitted. During Epochs II and III, however, the pulse period is well-constrained, resulting in distinctive pulse profiles, shown in the right column. Note that the scale of the y-axis in panel (c) is narrower than those of (f) and (i) in order to better illustrate the pulse profile during Epoch I.}
	\label{fig:Pdotsearches}
\end{figure*}

\quad
\begin{deluxetable*}{ccccc}
\tablecolumns{5}
\tabletypesize{\scriptsize}
\tablecaption{Results of the folding pulsation search for each epoch.\label{table:pulsedata}}
\tablehead{
\colhead{Epoch} 	& \colhead{$T_{ref}$ (MJD)}		& \colhead{$P$ (s)} 	& \colhead{$|\dot{P}|\ (10^{-8}\,\mathrm{s\,s^{-1}})$} 	& \colhead{Pulse Fraction (\%)}		
}
\startdata
I 					& $56113.28661210$				& \nodata 				& \nodata 												& $<4.5$					\\
II 					& $56113.92279551$				& $0.70121(20)$ 		& $<1.2$ 										& $21.5 \pm 1.5$ 		\\
III 				& $56145.10372569$ 				& $0.70117(9)$ 			& $<0.77$ 										& $40.9 \pm 0.5$ 		\\
\enddata
\end{deluxetable*}

We performed a timing analysis of both observations using the Stingray \citep{Huppenkothen2016Stingray:Software} and HENDRICS \citep{Bachetti} software packages in order to determine the pulse fraction, pulse period, and spin-up rate during each epoch. The results of this analysis are shown in Table \ref{table:pulsedata}. The pulse fraction, $PF$, is defined as follows
\begin{equation}
PF = \frac{F_{\rm max} - F_{\rm min}}{F_{\rm max} + F_{\rm min}}
\end{equation}
where $F_{\rm max}$ and $F_{\rm min}$ are the maximum and minimum fluxes in the pulse profile, respectively. All pulse fractions and corresponding errors quoted were calculated using a Monte Carlo analysis. Given a measured pulse period and derivative, we folded the observed events into a pulse profile with sixteen phase bins per cycle. The uncertainty in flux for each phase bin is given by a Poisson distribution. We sampled this distribution for each phase bin to produce a large number of simulated pulse profiles and passed these profiles through a Savitzky-Golay filter \citep{Savitsky1964}. We thus arrived at a distribution of smoothed profiles from which we extracted the mean pulse fraction and corresponding confidence regions. All uncertainties and upper limits quoted in this section and following sections correspond to 90\% confidence ranges unless otherwise indicated.

Before searching for and analyzing pulsations, we first determined the orbital phase of the observations. Using the orbital parameters reported by \cite{Falanga2015}, we determined the mid-eclipse times which occurred immediately before and after the observations according to the quadratic orbital change function
\begin{equation}
T_n = T_0 + nP_{\rm orb} + \frac{1}{2}n^2P_{\rm orb}\dot{P}_{\rm orb}
\end{equation}
where $T_0$ is the reference epoch (MJD 52846.6888), \textit{n} is the number of elapsed orbits, $P_{\rm orb}$ is the orbital period measured at $T_0$ (3.8919232\,days), and $\dot{P}_{\rm orb}$ is the time derivative of the orbital period measured at $T_0$ ($\mathrm{-3.77\times10^{-8}\,day\,day^{-1}}$). Defining the mid-eclipse times preceding and following each observation as orbital phases 0 and 1, respectively, we found that the first observation occurred between orbital phases $0.34$ and $0.61$, while the second observation occurred between phases $0.52$ and $0.61$. These orbital phases are determined to better than $10^{-5}$ and lie far from the eclipse ingress and egress times. Therefore we can be confident that there were no obscuration effects due to the supergiant companion.

We performed pulsation searches on the combined filtered and calibrated FPMA and FPMB events for each of the three epochs. When combining the FPMA and FPMB events for each epoch, we produced common good-time intervals (GTIs) in order to avoid introducing artificial variability due to non-simultaneous observation and differences in sensitivity between the two focal plane modules. We began our pulsation search by performing a dynamic search using the HENDRICS function \code{dyn\_folding\_search}. This function steps over time and pulse period, folds the events into a pulse profile with that period, and calculates the $Z_{4}^{2}$ statistic \citep{Buccheri1983}, a measure of the probability of pulsation detection, of the profile produced at each step. The probability density function of the $Z_{4}^{2}$ statistic is equivalent to that of a $\chi^{2}$ distribution with 8 degrees of freedom. Therefore, one can use the $\chi^{2}$ cumulative distribution function to determine the probability that a pulse profile with a given value of $Z_{4}^{2}$ has been produced by noise. For example, a profile with $Z_{4}^{2}=13$ has a 10\% probability of being produced by noise, therefore this can be considered a detection at 90\% confidence. A 5-sigma detection, corresponding to a probability of $5.7\times10^{-7}$ that a signal has been produced by noise, would yield a $Z_{4}^{2}$ statistic of 44.

The results of the dynamic pulsation search applied to Epoch III (Figure \ref{fig:Pdotsearches}g) confirm the presence of strong pulsations. The pulsations appear to remain persistent throughout the observation with a period around $0.701\,\mathrm{s}$ and without a large period derivative. The results of this test are less striking upon application to Epochs I and II (Figures \ref{fig:Pdotsearches}a and \ref{fig:Pdotsearches}d). The $Z_{4}^{2}$ statistic reaches only a fraction of the maximum value measured during Epoch III, and during Epoch I, there is no sign of pulsations. However, during Epoch II, pulsations appear to have begun with a period similar to that observed in Epoch III, reaching a maximum detection probability at the end of the observation.

We next simultaneously searched for the period and first period derivative of the pulsations for each of the three epochs using the HENDRICS function \code{folding\_search}. The results of this search are shown in the second column of Figure \ref{fig:Pdotsearches}. We were able to measure the pulse period and to put upper limits on the first derivative during Epochs II and III by fitting the resulting $Z_{4}^{2}$ distributions to 2-dimensional Gaussian distributions\footnote{The uncertainties reported for the pulse periods and the upper limits of the spin-up rates were determined using the widths of the fitted Gaussian distributions.} with mean pulse periods at $P_{\mathrm{II}}=0.70121(20)\,\mathrm{s}$ and $P_{\mathrm{III}}=0.70117(9)\,\mathrm{s}$, respectively. We have obtained upper limits on the instantaneous spin-up rates of $|\dot{P}_\mathrm{II}|<1.2\times10^{-8}\,\mathrm{s\,s^{-1}}$ and $|\dot{P}_\mathrm{III}|<7.7\times10^{-9}\,\mathrm{s\,s^{-1}}$. Note that there is a correlation between $P$ and $\dot{P}$ apparent in Figure \ref{fig:Pdotsearches}. This is not intrinsic to the source itself but is an artifact introduced by the search procedure. In addition to measuring the pulse periods and constraining the instantaneous spin-up rates, we have also placed an upper limit on the secular spin-up rate between Epoch II and Epoch III of $|\dot{P}_{\rm sec}|<10^{-10}\mathrm{\,s\,s^{-1}}$. After determining the pulse periods and the spin-up rates, we then folded the events into pulse profiles at the $Z_{4}^{2}$ maxima produced by the pulsation searches. These pulse profiles are shown in Figures \ref{fig:Pdotsearches}f and \ref{fig:Pdotsearches}i. We observe distinct pulsations in the pulse profiles of Epochs II and III. The probability that these profiles were produced by noise is vanishingly small, being less than $10^{-37}$ in both cases. 

In stark contrast, we were completely unable to constrain the pulse period during Epoch I. There are multiple local maxima of comparable amplitude in the $Z_{4}^{2}$ distribution. We therefore chose to fold the events into a pulse profile (see Figure \ref{fig:Pdotsearches}c) using the maximum nearest the values measured during Epochs II and III. This corresponds to a pulse period of $P_{\rm fold}=0.70113161\,\mathrm{s}$ and a first derivative of $\dot{P}_{\rm fold}=2.69\times10^{-9}\,\mathrm{s\,s^{-1}}$. The resulting profile has a pulse fraction of $<4.5\%$. This is relatively small compared to the pulse fractions of $21.5\% \pm 1.5\%$ during Epoch II and $40.9\% \pm 0.5\%$ during Epoch III. In addition to the small pulse fraction during Epoch I, the $Z_{4}^{2}$ value of the calculated pulse profile is less than 15 and corresponds to a probability of 7\% that the detection is due to noise. Furthermore, when the last 5000\,s of Epoch I are omitted from the pulsation search even this weak detection disappears, indicating that pulsations were absent until the very end of Epoch I. Therefore, we refer to Epoch I as the non-pulsing state.

The pulse periods that we have measured during Epochs II and III and the resulting pulse profiles are in line with previous measurements \citep[cf.][]{Moon2003, Naik2004, Raichur2010, Inam2010}. In particular, we have extrapolated previous results by applying an orthogonal distance regression to the pulse frequencies reported by \citeauthor{Inam2010}. We arrived at a spin up of $\dot{f}_{pulse}=2.589(8)\times10^{-11}\mathrm{\,Hz\,s}^{-1}$ during the interval 50093-52988\,MJD. When propagated forward to the beginning of Epoch III, a pulse period of $0.70093(2)\,\mathrm{s}$ is predicted. The discrepancy of $2.39(96)\times10^{-4}$\,s is small but nonzero. This is consistent with a piece-wise spin-up evolution, reported by \citeauthor{Inam2010}, in which the spin-up rate is variable. We also note that, although the pulse fraction increases with energy, the shapes of the pulse profiles during Epochs II and III do not appear to vary significantly with energy.

\section{Spectral Analysis}
\label{sec:spectral}

 \begin{deluxetable*}{ c c r r r }
 \tablecolumns{5}
 \tabletypesize{\scriptsize}
 \tablecaption{Values of spectral parameters determined by $\chi^2$ fitting of observed spectra. Two models are shown: a fully covered power law with a Fermi-Dirac-like cutoff modeled by \code{fdcut} (top), and a partially covered power law with an exponential cutoff, modeled by \code{cutoffpl} (bottom).\label{table:spectralPar}}
 \tablewidth{0pt}
 \tablehead{
 \colhead{Component} & \colhead{Parameter} 							& \colhead{Epoch I} 	& \colhead{Epoch II} 	& \colhead{Epoch III}
 }
 \startdata
 \noalign{\smallskip}
 \code{tbabs}						& $N_{\rm H}$ ($10^{22}\,\mathrm{cm}^{-2}$) 	& $16 \pm 5$ 				& $24^{+5}_{-4}$ 	            & $1.9^{+1.3}_{-0.9}$ 		\\
 \noalign{\smallskip}
 \hline
 \noalign{\smallskip}
 \multirow{4}{*}{\code{fdcut}} 		& $\Gamma$ 										& $1.0$\tablenotemark{$\dagger$} 	& $1.0$\tablenotemark{$\dagger$} 	& $1.0$\tablenotemark{$\dagger$}    \\
 \noalign{\smallskip}
 								 	& $E_{\rm cut}$ (keV) 							& $17.3^{+1.6}_{-2.3}$ 		        & $11.0^{+3.2}_{-5.1}$ 				& $9.1^{+2.5}_{-3.0}$ 		\\
 \noalign{\smallskip}
  								 	& $E_{\rm fold}$ (keV) 							& $6.7^{+0.8}_{-0.6}$   	        & $8.7^{+1.0}_{-0.8}$ 	            & $9.6 \pm 0.4$ 		    \\
 \noalign{\smallskip}
  								 	& Norm ($10^{-3}$) 								& $8.0^{+1.4}_{-0.9}$ 	            & $13.8^{+3.2}_{-2.5}$ 	            & $107^{+29}_{-16}$ 		\\
 \noalign{\smallskip}
 \hline
 \noalign{\smallskip}
 \multirow{3}{*}{\code{gauss}}		& $E_{\rm 6.4}$ (keV) 							& $6.36 \pm 0.04$ 		            & $6.36 \pm 0.06$ 			        & $6.51^{+0.09}_{-0.07}$	\\
 \noalign{\smallskip}
 			 						& $\sigma_{\rm 6.4}$ (keV) 						& $0.24 \pm 0.06$ 		            & $0.21\pm 0.10$ 	                & $0$\tablenotemark{$\dagger$}	    \\
 \noalign{\smallskip}
   								 	& Norm ($10^{-4}$) 								& $3.3 \pm 0.5$ 		            & $2.3^{+0.6}_{-0.5}$ 		    	& $1.9 \pm 0.7$		        \\
 \noalign{\smallskip}
 \hline
 \noalign{\smallskip}
 \multirow{3}{*}{\code{gauss}}		& $E_{\rm 13.5}$ (keV) 							& $13.5$\tablenotemark{$\dagger$} 	& $13.5$\tablenotemark{$\dagger$} 	& \multirow{3}{*}{\nodata}	\\
 \noalign{\smallskip}
 			 						& $\sigma_{\rm 13.5}$ (keV) 						& $2.2^{+0.8}_{-0.7}$ 	            & $1.7^{+1.8}_{-0.9}$ 		        & 					        \\
 \noalign{\smallskip}
   								 	& Norm ($10^{-4}$) 								& $2.3^{+1.6}_{-1.1}$ 	            & $1.1^{+1.7}_{-0.7}$		        & 					        \\
 \noalign{\smallskip}
 \hline
 \noalign{\smallskip}
 \multirow{2}{*}{\code{bbody}}		& $kT$ (keV) 									& $0.36^{+0.04}_{-0.06}$            & $0.31^{+0.03}_{-0.04}$            & $0.26 \pm 0.09$			\\
 \noalign{\smallskip}
   								 	& Norm ($10^{-3}$) 								& $3.6^{+3.5}_{-1.9}$	            & $19.6^{+24.6}_{-9.3}$ 			& $56^{+195}_{-45}$		    \\
 \noalign{\smallskip}
 \hline
 \noalign{\smallskip}
 \multirow{2}{*}{\code{bbody}}		& $kT$ (keV) 									& \multirow{2}{*}{\nodata}          & \multirow{2}{*}{\nodata}          & $1.46 \pm 0.07$		    \\
 \noalign{\smallskip}
   								 	& Norm ($10^{-3}$) 								&  						            &  					                & $2.3 \pm 0.4$ 		    \\
 \noalign{\smallskip}
 \hline
  \noalign{\smallskip}
 									& Absorbed Flux\tablenotemark{$a$} ($10^{-11}\,\mathrm{erg}\,\mathrm{cm}^{-2}\,\mathrm{s}^{-1}$)	& $6.26^{+0.15}_{-0.32}$	& $6.67^{+0.06}_{-0.49}$	&	$99.3^{+0.8}_{-15.4}$	\\
 \noalign{\smallskip}
 									& Unabsorbed Flux\tablenotemark{$a$} ($10^{-10}\,\mathrm{erg}\,\mathrm{cm}^{-2}\,\mathrm{s}^{-1}$)	& $1.46^{+0.56}_{-0.33}$	& $2.82^{+1.16}_{-0.69}$	&	$11.6^{+4.0}_{-0.9}$	\\
 \noalign{\smallskip}
 									& Unabsorbed Luminosity ($10^{37}\,\mathrm{erg}\,\mathrm{s}^{-1}$)	& $6.43^{+2.48}_{-1.44}$	& $12.4^{+5.1}_{-3.0}$	&	$51.0^{+17.7}_{-4.1}$	\\
 \noalign{\smallskip}
 									& $\chi^2/\mathrm{d.o.f.}$						&  693/630	(1.10)		& 681/679 (1.00)		& 1091/1019 (1.07)			\\
 \noalign{\smallskip}
 \hline
 \hline
 \noalign{\smallskip}
 \multirow{2}{*}{\code{tbpcf}}		& $N_{\rm H}$ ($10^{23}\,\mathrm{cm}^{-2}$) 	& $7.8^{+1.9}_{-2.1}$ 	            & $5.6^{+0.9}_{-1.2}$ 		        & $26^{+18}_{-10}$ 	    \\
 \noalign{\smallskip}
									& $f_{\rm covering}$ (\%) 						& $51 ^{+7}_{-8}$ 		            & $61^{+2}_{-4}$ 		            & $15^{+7}_{-6}$ 			\\
 \noalign{\smallskip}
 \hline
 \noalign{\smallskip}
 \multirow{3}{*}{\code{cutoffpl}} 	& $\Gamma$ 										& $0.5$\tablenotemark{$\dagger$}    & $0.5$\tablenotemark{$\dagger$}    & $0.5$\tablenotemark{$\dagger$} 		\\
 \noalign{\smallskip}
 								 	& $E_{\rm cut}$ (keV) 							& $8.4 \pm 0.6$ 	                & $8.9^{+0.4}_{-0.3}$ 			    & $9.1 \pm 0.2$ 		\\
 \noalign{\smallskip}
 								 	& Norm ($10^{-3}$) 								& $6.7^{+1.5}_{-1.2}$ 	            & $7.5^{+0.7}_{-0.9}$ 	            & $53 \pm 4$ 		    \\
 \noalign{\smallskip}
 \hline
 \noalign{\smallskip}
 \multirow{3}{*}{\code{gauss}}		& $E_{\rm 6.4}$ (keV) 							& $6.35^{+0.04}_{-0.05}$             & $6.34^{+0.06}_{-0.07}$            & $6.52 \pm 0.08$           \\
 \noalign{\smallskip}
  			 						& $\sigma_{\rm 6.4}$ (keV) 						& $0.22 \pm 0.07$	                & $0.24^{+0.10}_{-0.09}$ 	        & $0$\tablenotemark{$\dagger$}  \\
 \noalign{\smallskip}
  								 	& Norm ($10^{-4}$) 								& $3.6^{+0.7}_{-0.6}$ 		        & $2.6^{+0.7}_{-0.5}$ 		        & $2.0^{+0.8}_{-0.9}$ 		    \\
 \noalign{\smallskip}
 \hline
 \noalign{\smallskip}
 \multirow{3}{*}{\code{gauss}}		& $E_{\rm 13.5}$ (keV) 							& $13.5$\tablenotemark{$\dagger$} 	& $13.5$\tablenotemark{$\dagger$}   & \multirow{3}{*}{\nodata}	\\
 \noalign{\smallskip}
 			 						& $\sigma_{\rm 13.5}$ (keV) 						& $4.9^{+1.3}_{-1.1}$            	& $1.7^{+3.3}_{-1.1}$                & 			                \\
 \noalign{\smallskip}
   								 	& Norm ($10^{-4}$) 								& $7.8 \pm 3.1$ 	                & $1.0^{+3.0}_{-0.7}$ 	            &  	                        \\
 \noalign{\smallskip}
 \hline
 \noalign{\smallskip}
 \multirow{2}{*}{\code{bbody}} 		& $kT$ (keV) 									& \multirow{2}{*}{\nodata} 	        & \multirow{2}{*}{\nodata}	        & $0.23^{+0.08}_{-0.07}$	\\
 \noalign{\smallskip}
    								& Norm ($10^{-1}$) 								&  							        &  							        & $1.1^{+72.3}_{-1.0}$ 		\\
 \noalign{\smallskip}
 \hline
 \noalign{\smallskip}
 \multirow{2}{*}{\code{bbody}} 	    & $kT$ (keV) 									& \multirow{2}{*}{\nodata} 	        & \multirow{2}{*}{\nodata}	        & $1.46^{+0.08}_{-0.09}$		\\
 \noalign{\smallskip}
    								& Norm ($10^{-3}$) 								&  							        &  							        & $3.0^{+0.8}_{-0.7}$ 		\\
 \noalign{\smallskip}
 \hline
 \noalign{\smallskip}
 									& Absorbed Flux\tablenotemark{$a$} ($10^{-11}\,\mathrm{erg}\,\mathrm{cm}^{-2}\,\mathrm{s}^{-1}$)	& $6.49^{+0.06}_{-0.24}$	& $6.90^{+0.08}_{-0.09}$	&	$105^{+2}_{-18}$	\\
 \noalign{\smallskip}
 									& Unabsorbed Flux\tablenotemark{$a$} ($10^{-10}\,\mathrm{erg}\,\mathrm{cm}^{-2}\,\mathrm{s}^{-1}$)	& $1.06^{+0.17}_{-0.14}$	& $1.18^{+0.08}_{-0.11}$	&	$12.7^{+1.1}_{-0.6}$	\\
 \noalign{\smallskip}
 									& Unabsorbed Luminosity ($10^{37}\,\mathrm{erg}\,\mathrm{s}^{-1}$)	& $4.65^{+0.73}_{-0.62}$	& $5.20^{+0.34}_{-0.47}$	&	$55.8^{+4.8}_{-2.8}$	\\
 \noalign{\smallskip}
 									& $\chi^2/\mathrm{d.o.f.}$						&  705/632	(1.11)		& 681/681 (1.00)		& 1090/1019 (1.07)			\\
 \noalign{\smallskip}
 \enddata
 \tablenotetext{a}{Fluxes are reported for FPMA in the energy range $2-10$ keV.}
 \tablenotetext{\dagger}{Values marked with a dagger were frozen during fitting and therefore have no error estimates.}
 \end{deluxetable*}

We also performed a spectral analysis of each of the three epochs using Xspec \citep[v.12.10.0][]{Arnaud1996}. We simultaneously fit the spectra measured by FPMA and FPMB while including a relative constant to account for small ($<10\%$) differences in flux between the two focal plane modules. In addition, for all spectral models described in this section, we have included an absorber in the form of the \code{tbabs} component. This component compensates for absorption due to Galactic material. We fixed the equivalent HI column density of this component at $N_{\rm H} = 4.58\times10^{21}\ \mathrm{cm}^{-2}$, determined from the full-sky HI survey, HI4PI \citep{BenBekhti2016}. The spectral fits were performed using interstellar medium abundances reported by \cite{Wilms2000}. 

\begin{figure}[t!]
\gridline{\fig{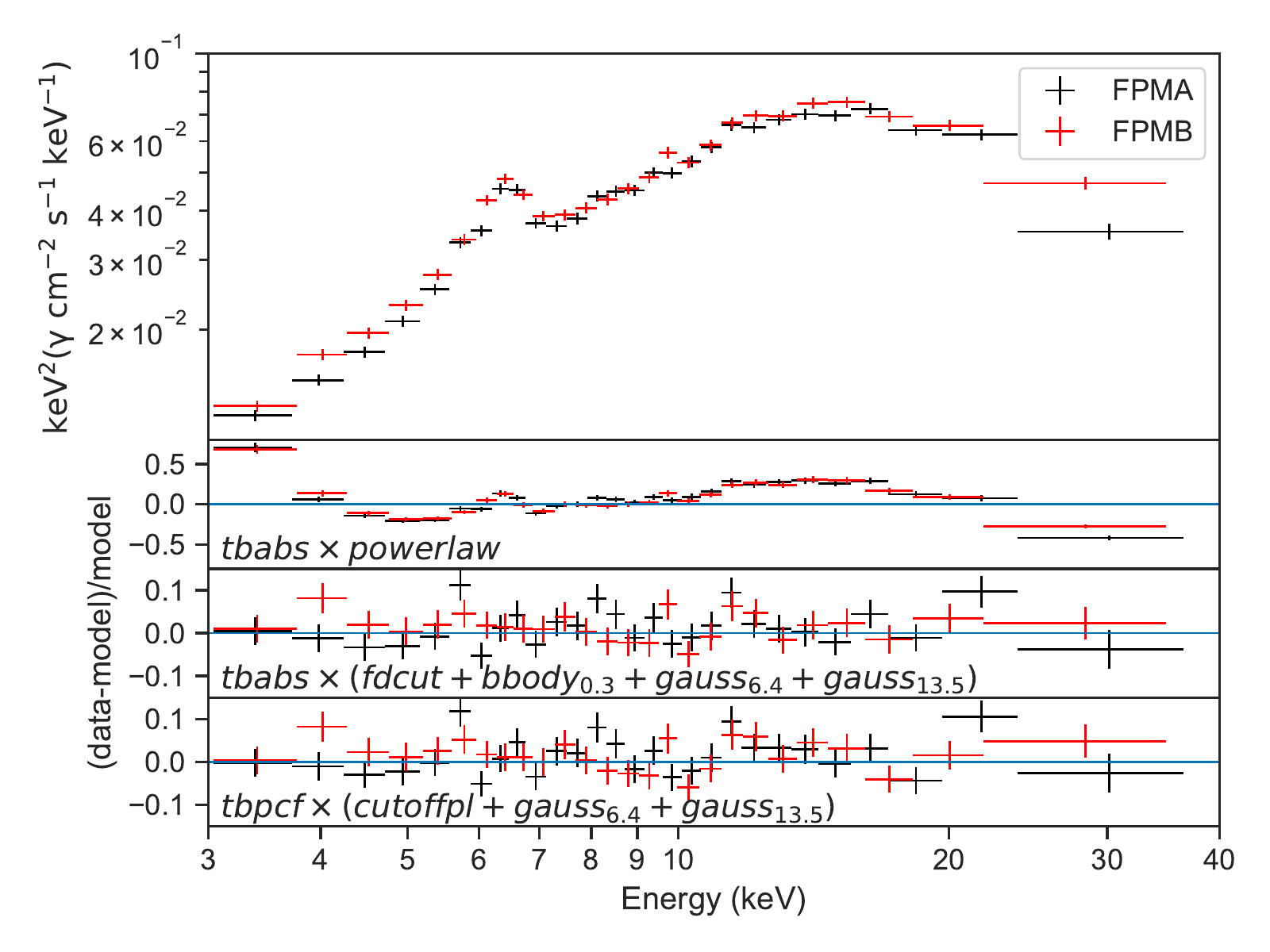}{0.375\textwidth}{(a) Epoch I}}
\gridline{\fig{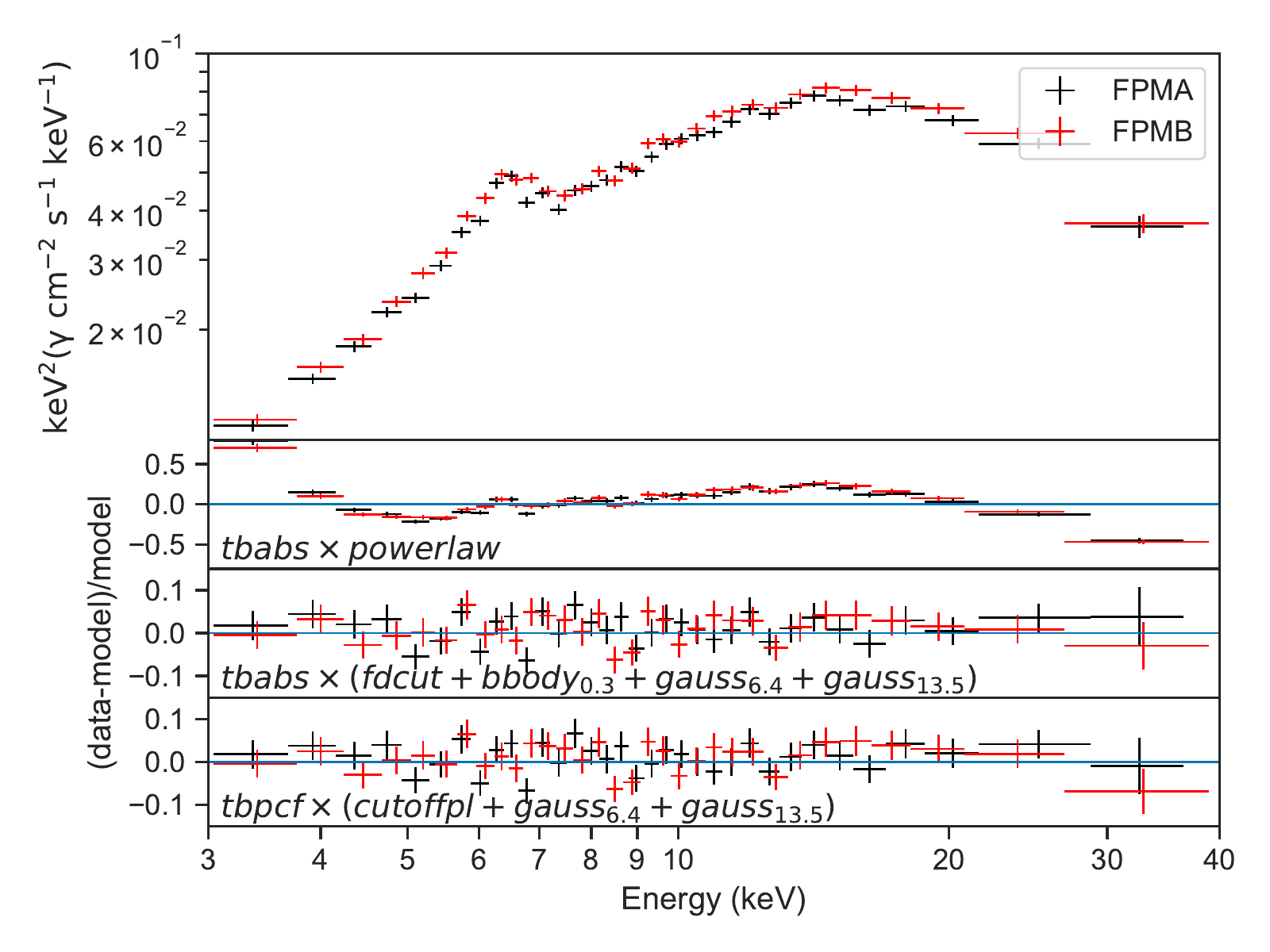}{0.375\textwidth}{(b) Epoch II}}
\gridline{\fig{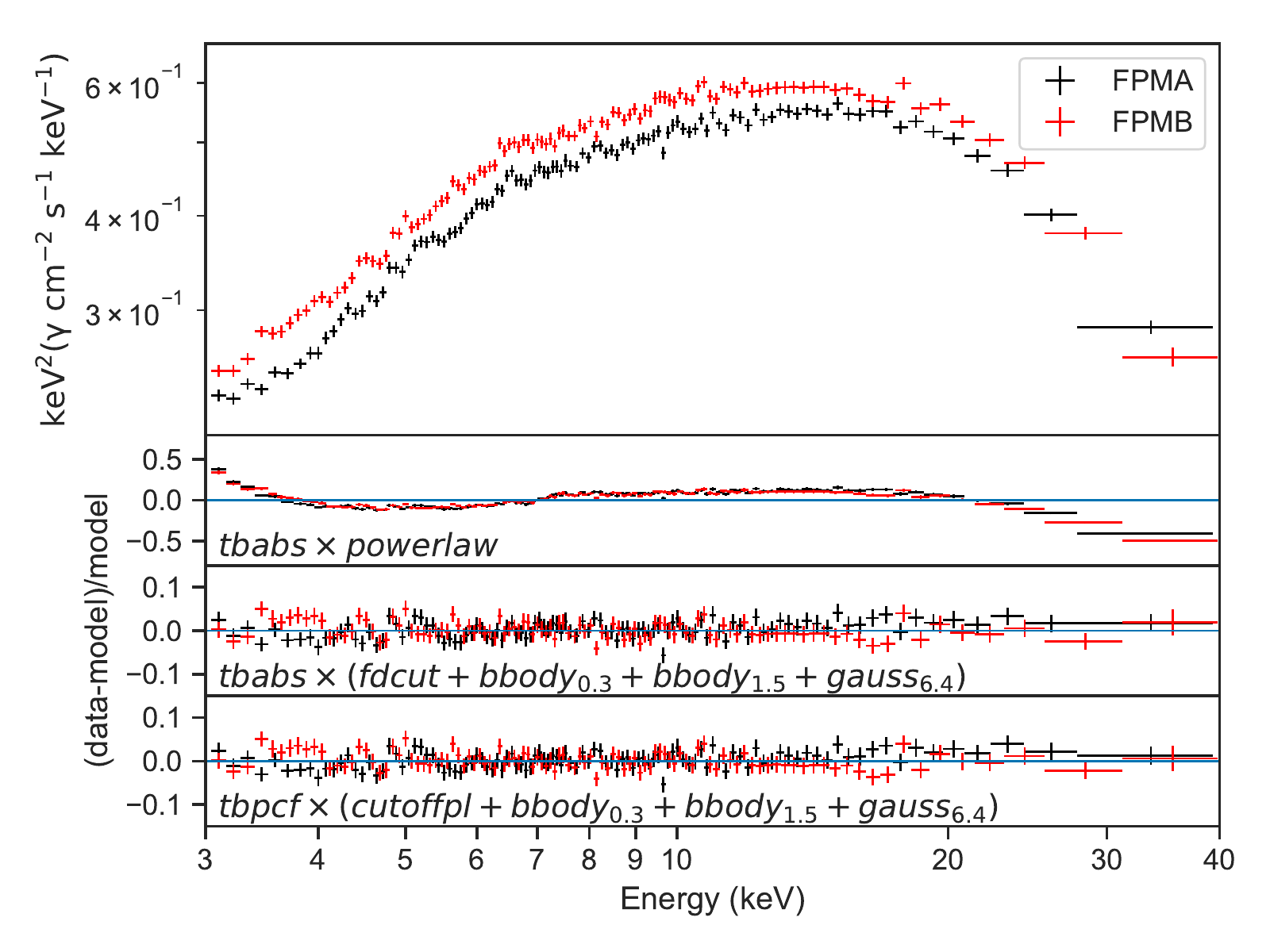}{0.375\textwidth}{(c) Epoch III}}
\caption{Observed spectra for each epoch are shown unfolded against a model with constant (energy-independent) flux in the top panels, and fitting residuals for three different models are shown in the lower panels. The FPMA spectra are shown in black while the FPMB spectra are shown in red. The data is consistently described well by a partially absorbed power law with a high energy cutoff. While the shape of the continuum remains relatively constant, the covering fraction and absorbing column vary between successive epochs, with the covering fraction decreasing significantly between Epoch II and Epoch III.}
\label{fig:spectra}
\end{figure}

The spectra for each epoch are shown in Figure \ref{fig:spectra}, and the results of our spectral analysis are presented in Table \ref{table:spectralPar}. Each panel in Figure \ref{fig:spectra} also includes residuals for three different models, including a simple absorbed power law meant to illustrate additional structure in the spectra. We have found two models that provide fits of similar quality and which result in physically reasonable parameters. Motivated by previous work by, e.g., \citet{Woo1995WindObservations}, \citet{Angelini1991THEL}, and \citet{2014HEAD...1411702P}, the first model we investigated was an absorbed power law with a phenomenological cutoff, named \code{fdcut} \citep{Tanaka1986ObservationsSources} for its resemblance to the Fermi-Dirac distribution, which has both a cutoff energy and folding energy and can be written
\begin{equation}
f_{\rm FD}(E) = \frac{E^{-\Gamma}}{1 + e^{(E-E_{\rm cut})/E_{\rm fold}}}
\end{equation}
where $\Gamma$ is the photon index, $E_{\rm cut}$ is the cutoff energy, and $E_{\rm fold}$ is the folding energy. The absorber in this model is fully covering and is modeled by \code{tbabs} \citep{Wilms2000}. The second model consists of a power law with an exponential cutoff, represented by the Xspec model \code{cutoffpl}, partially covered by an absorber modeled by \code{tbpcf}. In addition to these base models, we found that the fits benefited from the addition of secondary components, differing depending on the epoch. Below, we describe each of these models in more detail. 

In order to compare the usefulness of additional model components, we use the Bayesian Information Criterion \citep[BIC;][]{Schwarz1978EstimatingModel}. In the case of $\chi^2$ fitting in Xspec, the BIC is given by
\begin{equation}
\mathrm{BIC} = k\ln(n) + \chi^2
\end{equation}
where $n$ is the number of PHA bins being fitted and $k$ is the number of parameters estimated by a given model. For a given data set, model selection can be achieved by minimizing the BIC, which penalizes models with many parameters. For our analysis, $n$ lies between 600 and 1100 bins, meaning that removing one parameter from a model without a change in $\chi^2$ results in a decrease in the BIC of $\Delta \mathrm{BIC} \approx -7$. In determining the impact of adding or subtracting components, this may be considered one ``unit'' of model improvement.

The spectra observed during the two observations are qualitatively different, as is visible in Figure \ref{fig:spectra}. For Epochs I and II, an absorbed \code{fdcut} model alone results in significant excess residuals below 4\,keV, around 6.4\,keV, and above 10\,keV. The excess around 6.4\,keV is consistent with previous detections of an Fe K$\alpha$ line in \smc, such as those by \citet{Woo1995ROSATState} and \citet{Naik2004}. We included a Gaussian component at this energy to model the line, allowing both the position and width of the line to vary. To address the low-energy excess, we added a black body component with temperature $kT_{\rm BB}<0.5$\,keV. Such a component has previously been detected in observations of \smc\ by the {\em Chandra X-ray Observatory} and {\em XMM-Newton} \citep{Neilsen2004PHASEX-1, Hickox2005}. Each of these components decreases the BIC by about 100, and adding both of these components results in a combined improvement to the fit of $\Delta \mathrm{BIC} = -304$ for Epoch I and $\Delta \mathrm{BIC} = -214$ for Epoch II, indicating that the improvement is significant. Adding these components does not resolve the ``bump" above 10\,keV. This residual resembles the 10\,keV feature observed in other accreting pulsars \citep[e.g.][]{Coburn2002MagneticExplorer, Mihara1995ObservationalGinga, Santangelo1998BeppoSAXX-3}, leading us to include a Gaussian component at $E=13.5$\,keV. We froze the position of this component in order to better constrain other parameters, while the width of the Gaussian was allowed to vary. Adding this component does not result in a significant improvement to the fit, with $\Delta \mathrm{BIC} = -7$ for Epoch I and $\Delta \mathrm{BIC} = +6$ for Epoch II. However, because the residuals are clearly reduced, and this feature is consistent with previous studies of accreting pulsars, the component is included in the final fit.

In the case of Epoch III, the \code{fdcut} model again requires a low-temperature ($kT_{\rm BB}<0.5$\,keV) black body in order to explain excess flux below 4\,keV, the addition of which improves the fit by $\Delta \mathrm{BIC} = -93$. In addition, although the excess near 6.4\,keV is not as prominent as in the previous two Epochs, adding a line at this energy improves the fit significantly ($\Delta \mathrm{BIC} = -46$). However, the width of the line is poorly constrained, leading us to freeze it at $\sigma = 0$. We also found that for Epoch III, adding a $kT\approx1.5$\,keV blackbody component, like the one included by \citet{2014HEAD...1411702P} in their analysis of this observation, improves the fit by $\Delta \mathrm{BIC} = -46$ while also eliminating an excess of flux above 20\,keV. Similar blackbody components with temperatures ranging between 1.2\,keV and 3\,keV have proved useful for modeling the spectra of several BeXRBs \citep[e.g.][]{Reig1999, LaPalombara2009XMM-NewtonRXJ0440.9+4431, Caballero2013ASuzaku}. In contrast to Epochs I and II, the 13.5\,keV bump is not observed during Epoch III, and its addition to the \code{fdcut} model does not improve the fit nor can this component be easily constrained.

The partially covered cutoff power law provides a similarly good fit to the data as the Fermi-Dirac-like model. However, the secondary components differ somewhat. Adding a line at $E=6.4$\,keV to the base model again improves the fit significantly during Epochs I and II, with $\Delta \mathrm{BIC} = -97$ and $\Delta \mathrm{BIC} = -43$, respectively. Adding this line to Epoch III does not result in a striking improvement (only $\Delta \mathrm{BIC} = -6$), but the position of the Gaussian is constrained to the same value as in the \code{fdcut} model. The 13.5\,keV bump is also added to this model for Epochs I and II, again slightly improving the fits. Unlike the absorbed \code{fdcut} model, the partially covered cutoff power law does not require a low-temperature black body to resolve excess emission below 4\,keV during Epochs I and II. This component, along with the $kT_{\rm BB}\approx1.5$\,keV black body, remains in Epoch III. Adding each of these black body components individually yields different results. Including only the warm $kT_{\rm BB}\approx1.5$\,keV black body improves the fit by $\Delta \mathrm{BIC} = -142$, while adding only the low-temperature $kT_{\rm BB}<0.5$\,keV black body does not improve the fit, yielding $\Delta \mathrm{BIC} = +6$. However, when both components are included, the fit is improved by $\Delta \mathrm{BIC} = -170$. In other words the combination of the low-temperature and high-temperature black body components improves the fit more than each of these components individually.

None of the three spectra can be fit to a simple one-component model, instead requiring several secondary components in order to properly fit the \nustar\ observations. In order to reduce degeneracies resulting from the number of parameters used in the final models, we froze some key parameters at values which are consistent with initial estimates. As mentioned above, the position of the bump above 10\,keV was frozen at $E=13.5$\,keV, and the width of the Fe K$\alpha$-like line in Epoch III was frozen at $\sigma=0$\,keV. In addition to these, we froze the photon indices across all three Epochs at $\Gamma=1.0$ for the \code{fdcut} model and $\Gamma=0.5$ for the \code{cutoffpl} model.

The final model components, parameter estimates, and fit information are shown in Table \ref{table:spectralPar}. Here we remind the reader that the uncertainties quoted on spectral parameters represent 90\% confidence regions. We found that for both models, the absorption parameters vary between epochs, while the underlying power law models show less variability. In the case of the \code{fdcut} model, the absorbing column density decreases by an order of magnitude from $N_{\rm H} = (2.4^{+0.5}_{-0.4})\times10^{23}\,\mathrm{cm}^{-2}$ during Epoch II, to $N_{\rm H} = (1.9^{+1.3}_{-0.9})\times10^{22}\,\mathrm{cm}^{-2}$ during Epoch III. As we have shown, the pulse fraction simultaneously increases between these two epochs. The shape of the \code{fdcut} component on the other hand remains consistent between Epochs II and III, with the cutoff energy of $E_{\rm cut}\approx 10\,{\rm keV}$ folding energy of $E_{\rm fold}\approx 9\,{\rm keV}$. However, Epoch I has a slightly higher cutoff energy and lower folding energy: $E_{\rm cut} = 17.3^{+1.6}_{-2.3}\,{\rm keV}$ and $E_{\rm fold} = 6.7^{+0.8}_{-0.6}\,{\rm keV}$.

On the other hand, in the case of the partially covered cutoff power law, the shape of the \code{cutoffpl} component stays constant. The exponential cutoff is consistent with $E_{\rm cut}\approx 9\,{\rm keV}$ during all three epochs. The absorption parameters show little variation between Epoch I and Epoch II, but the covering fraction drops by a factor of four from $f_{\rm covering} = (61^{+2}_{-4})\%$ in Epoch II to $f_{\rm covering} = (15^{+7}_{-6})\%$ in Epoch III. Between these epochs, the column density appears to increase by a factor of a few, but this parameter is poorly constrained during Epoch III due to the low covering fraction.

In both models, the underlying continuum increases in flux between successive epochs while the flux of the apparent Fe K$\alpha$ line remains constant. Taken together, these observations indicate that the increase in total flux between epochs cannot be attributed solely to the absorption included in the models described above, and that the source of the Fe K$\alpha$ line is likely distinct from the source of the continuum (e.g., originating in the photoionization region surrounding the central X-ray source). In addition, the appearance of the $kT_{\rm BB}\approx1.5\,{\rm keV}$ black body in Epoch III, observed in both the \code{fdcut} model and the \code{cutoffpl} model, may indicate that the emitting region responsible for this component either did not exist or was obscured during Epochs I and II.

\section{Discussion}
\label{sec:discussion}


Our timing analysis has shown that the source was observed in a non-pulsing state during Epoch I which subsequently evolved into a pulsing state, observed in Epoch II. During Epoch III, about a month after Epoch II, the pulsations had increased in strength, with the pulse fraction increasing by nearly a factor of two. At the same time, our spectral analysis has shown that for all three epochs, the emission of the source can be described by two different models: a fully covered power law with a phenomenological Fermi-Dirac-like cutoff, and a partially absorbed power law with an exponential cutoff. Each of these models requires additional components, but we found that both models are consistent with variable absorption parameters between the low and high pulse fraction states. In particular, the Fermi-Dirac model exhibits a decrease in absorption column density between Epoch II and Epoch III, and the cutoff power law is consistent with a decrease in the covering fraction (and a poorly constrained increase in column density), between the low and high pulse fraction states. In addition, the luminosity was observed to gradually increase between the non-pulsing Epoch I and the pulsing Epoch II.

In order to synthesize these results, we propose that the pulsing region was observed emerging from behind absorbing material. Given that Epochs I and II took place near the end of the low state, as illustrated in Figure \ref{fig:BATcurve} and that the super-orbital period has been attributed to a warped precessing accretion disk \citep{Wojdowski1998, Clarkson2003, Dage2018}, the absorbing material obscuring the pulsing region is likely part of the accretion disk. In short, the warped accretion disk absorbs and scatters the pulsed emission from the neutron star, leading to the absence of detected pulsations in the low state and a gradual turn-on of pulsations as the disk moves out of the line of sight. This picture is consistent with the opaque inner disk region described by \citet{Hickox2005} to explain apparent reprocessing of pulsed emission. Their analysis describes the case when both the neutron star and the inner regions of the warped accretion disk are visible to the observer, while ours describes the opposite case when the warped disk lies between the neutron star and the observer, obscuring the pulsing emission regions.

The relatively high absorption column density of the partially covered cutoff power law in Epoch III does not immediately fit within this interpretation. Although this column density is not particularly well constrained, it is still well above the values measured during the first two epochs. This increased column density is accompanied by an increase in the brightness of the power law itself; in other words, the increased flux between the two observations cannot be attributed solely to the absorption included in the models presented here. Thus one interpretation of the combination of a relatively high column density and a relatively low covering fraction is that much of the absorber is completely Compton thick during Epochs I and II. The partially covering absorber represented by \code{tbpcf}, then, only models the optically thinner regions of the accretion disk leading the covering fraction and column density to be underestimated for the first two epochs. During Epoch III, according to this interpretation, the source is observed through an overall less opaque region of the accretion disk so that a higher column density is measurable.

Pulse fraction variability, including pulse drop-out, has been observed in several other accreting pulsars. In some cases, this variability has been attributed to changes in accretion via the propeller effect \citep{Illarionov1975}. These include HMXBs Vela~X-1 and GX~301$-$2, which have been shown to exhibit off-states during which the sources drop in luminosity and pulsations are no longer detected \citep{Kreykenbohm2008HighStates, Furst2011AstronomyXMM-Newton}. LMC~X-4, in which pulse drop-out and turn-on have been observed during the high state \citep{Brumback2018DiscoveryX-4}, presents a different case. Still others, such as the low mass X-ray binary Her~X-1, exhibit pulse fraction variability attributed to obscuration by warped accretion disks \citep{Kuster2002}. Of these examples, the case of variable obscuration in Her~X-1 is most analogous to the behavior we have observed in \smc.

\section{Conclusions}
\label{sec:conclusions}


We have performed spectral and timing analyses of the accreting neutron star binary \smc\ for three separate epochs occurring during two \nustar\ observations. Our timing analysis confirmed that the source was observed in the midst of a turn-on of pulsations, which subsequently increased in strength before strong pulsations were observed a month later. Our spectral analysis, which showed variable absorber parameters and luminosity, led us to conclude that the non-pulsing state was due to obscuration of the pulsing region by a warped accretion disk, and that the gradual turn-on was due to the emergence of the pulsing emission from behind the disk.

Similarly to \smc, ULXPs are also known to exhibit variability in their luminosities and pulse fractions. In particular, the gradual change in pulse fraction observed in the beginning of the 2014 observation of M82~X-2 \citep{Bachetti2014} may share the same physical origin as the pulse fraction variability we have observed in \smc. In that case, the super-orbital periods observed in ULXPs may be attributable to precessing accretion disks which periodically obscure the pulsing source, resulting in variability in the observed pulse fractions. Spectral and timing analyses at different points in the super-orbital cycles of known ULXPs, like the analysis we have carried out for \smc, may help to illuminate the accretion mechanism and causes of variability in this recently discovered class of X-ray binary.

\acknowledgements
This work was supported under NASA grant No. NNG08FD60C, and made use of data obtained with \nustar, a project led by Caltech, funded by NASA and managed by NASA/JPL and has utilized the NUSTARDAS software package, jointly developed by the ASDC (Italy) and Caltech (USA). We would also like to thank the anonymous referee for providing helpful comments that helped to improve the quality of this paper.

\software{Stingray \citep{Huppenkothen2016Stingray:Software}, HENDRICS \citep{Bachetti}, NUSTARDAS, MaLTPyNT \citep{Bachetti}, DS9 \citep{ds9}}

\end{document}